\documentclass[opre]{informs3nothing}  

\DoubleSpacedXI 

\usepackage{latexsym, bbm, enumerate, amssymb, amsmath, libertine,tikz}
\usepackage{gensymb}
\usepackage{booktabs}
\usepackage{float}
\usepackage[utf8]{inputenc}
\usepackage[T1]{fontenc}
\usepackage{subcaption}
\usepackage{multirow}
\usepackage{array}
\usepackage{colortbl}

\usepackage{natbib}
\usepackage{hyperref}
\bibpunct[, ]{(}{)}{,}{a}{}{,}%
\def\bibfont{\small}%
\graphicspath{{./figures/}}

\TheoremsNumberedThrough     
\ECRepeatTheorems

\EquationsNumberedThrough    

\newcommand{\paranth}[1]{\left(#1\right)}

\newcommand{\curly}[1]{\left\{#1\right\}}

\TITLE{A Bayesian framework for opinion dynamics models}
\RUNTITLE{A Bayesian framework for opinion dynamics models}

\RUNAUTHOR{Chen and Zaman}

\ARTICLEAUTHORS{%
\AUTHOR{Yen-Shao Chen}
\AFF{School of Management, Yale University, New Haven, CT 06511, \EMAIL{yen-shao.chen@yale.edu}} 
\AUTHOR{Tauhid Zaman}
\AFF{School of Management, Yale University, New Haven, CT 06511, \EMAIL{tauhid.zaman@yale.edu}}
} 

\ABSTRACT{
This work introduces a subjective Bayesian framework for the individual update rules used in opinion dynamics models. An individual's initial opinion is represented by the center of a prior belief about an unknown state, and the updated opinion by the posterior mean. After observing a signal, the individual interprets it through a subjective likelihood, which may incorporate perceived bias and noise, and updates the belief by Bayes' rule. Varying the prior and perceived-signal distributions generates four principal response classes: linear updating, saturation, tail rejection, and signal tracking. A sufficiently separated bimodal prior generates local overreaction, mixture signals generate localized attenuation through source attribution, and perceived signal bias generates directional reversal over a finite region. The framework provides Bayesian microfoundations for established response functions and shows that updates often viewed as irrational can be Bayes-consistent under a receiver's subjective beliefs.
}


\begin{document}
\maketitle
\section{Introduction}
Opinion dynamics, the study of how individuals form and change their opinions, draws on psychology, sociology, economics, mathematics, and computer science. It informs research on political polarization, echo chambers, consensus formation, and information cascades in social networks. Although the field has developed many useful models (see Table~\ref{table:model_psychology_mapping}), these models use different mathematical formulations and often make different implicit assumptions about how people process information.

A central challenge is to reconcile observed behavior with a coherent account of belief updating. Anchoring, confirmation bias, backfire effects, and overreaction can all depart from the predictions of simple models of rational information processing. Opinion dynamics models capture these patterns effectively, but often introduce them as separate modifications to a baseline rule. The resulting models are useful for particular phenomena but can be difficult to compare within a common theoretical framework.

\subsection{Our Contributions}
This paper argues that many opinion-update patterns that appear inconsistent with simple linear averaging can emerge from Bayesian updating under the receiver's subjective prior and perceived-signal model. The resulting updates are Bayes-consistent with that subjective model; they need not coincide with inference under the objective signal-generating process.

Our contributions are threefold. First, we provide a receiver-side Bayesian microfoundation for one-time opinion updates. The framework begins with an unknown state $\Theta$. An individual holds a prior distribution over $\Theta$ and, after observing a signal realization $x$, interprets it through a subjective likelihood $l_{X|\Theta}(x|\theta)$. The shift from the prior center to the posterior mean defines the \textit{opinion shift function}. This formulation separates the receiver's updating problem from strategic information design and makes the behavioral assumptions explicit in the prior and perceived-signal distributions.

\begin{figure}[htbp]
    \centering
    \includegraphics[scale=0.2]{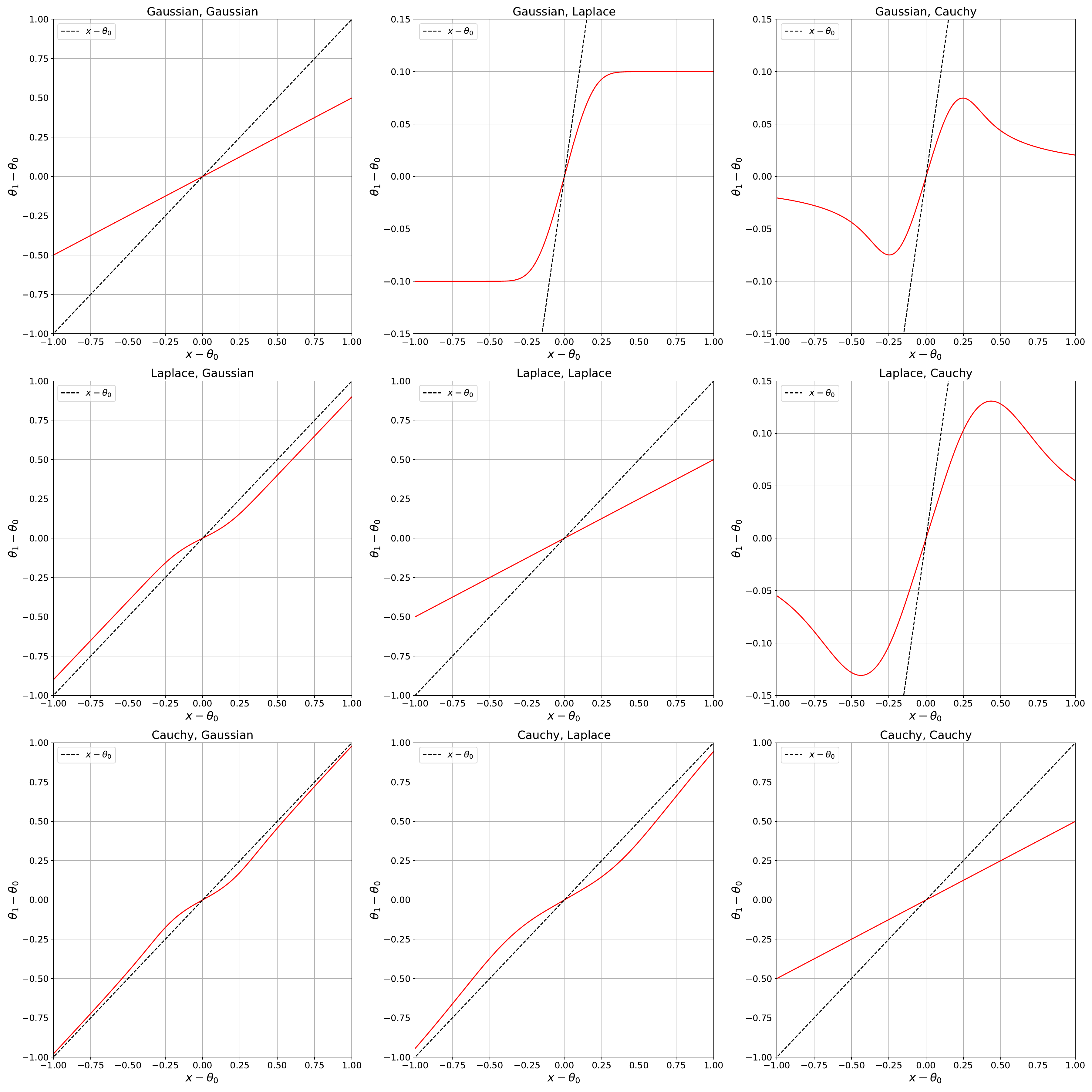}
    \caption{Opinion shift as a function of signal deviation for several prior and likelihood combinations with $\sigma_0=\sigma_\epsilon$.}
    \label{fig:shift_grid_equal}
\end{figure}

Second, we derive a distributional taxonomy of response functions. Gaussian--Gaussian, equal-scale Laplace--Laplace, and Cauchy--Cauchy pairs produce exact linear rules (Theorems~\ref{thm:gaussian_degroot},~\ref{thm:laplace_laplace_equal_scale}, and~\ref{thm:cauchy_degroot}). Other prior--likelihood combinations produce saturation (Theorems~\ref{thm:laplace_laplace_unequal} and~\ref{thm:gaussian_laplace}), tail rejection (Theorems~\ref{thm:gaussian_cauchy} and~\ref{thm:laplace_cauchy}), or signal tracking (Theorem~\ref{thm:laplace_laplace_unequal} and Subsection~\ref{sec:overreaction}). Near the prior opinion, the update is locally linear; under a concentrated prior, the likelihood score diagnoses its response shape, while relative tail behavior determines the large-signal class for the distributions studied here. Table~\ref{table:models} and Figure~\ref{fig:shift_grid_equal} summarize these correspondences.

Third, the framework identifies localized mechanisms beyond the four large-signal classes. A sufficiently separated bimodal prior generates local overreaction, mixture signals generate localized attenuation through source attribution (Theorem~\ref{thm:mixture_signal}), and perceived signal bias generates directional reversal and hence a backfire effect over a finite region. These results illustrate how new response rules, including bounded shifts, can be constructed from previously unexplored prior--likelihood combinations without adding ad hoc adjustments to the posterior update.

\begin{table}[htbp]
\centering
\caption{Key opinion dynamics models and their common behavioral interpretations. The interpretations motivate the response shapes but need not be their unique psychological mechanisms.}
\label{table:model_psychology_mapping}
\begin{tabular}{p{0.45\textwidth}|p{0.45\textwidth}} 
\hline
\textbf{Opinion Dynamics Model} & \textbf{Common Behavioral Interpretation} \\
\hline
DeGroot Model \citep{degroot1974reaching}& Social influence and conformity \\
\hline
Friedkin--Johnsen Model \citep{friedkin_social_1990} & Belief perseverance, anchoring effects \\
\hline
Bounded Confidence Model \citep{deffuant_mixing_2001, hegselmann_opinion_2002} & Confirmation bias \\
\hline
Bounded Shift Model\textsuperscript{*} & Anchoring and distance effects\\
\hline
Underreaction Model \citep{barberis_model_1998, dandekar_biased_2013} & Anchoring effect, conservatism\\
\hline
Overreaction Model \citep{barberis_model_1998} & Overreaction to information, representativeness \\
\hline
Backfire Model \citep{jager_uniformity_2005} & Backfire effect, reactance \\
\hline
Bayesian Opinion-Update Model \citep{martins_bayesian_2009} & Belief revision under a subjective probabilistic model \\
\hline
\multicolumn{2}{l}{\textsuperscript{*} This paper.}
\end{tabular}
\end{table}

\newlength{\taxonomycellwidth}
\begin{table}[htbp]
    \centering
    \settowidth{\taxonomycellwidth}{Saturating (Bounded Shift)}
    \begin{tabular}{|c|c|c|c|}
    \hline
    \textbf{} & \textbf{Gaussian Signal} & \textbf{Laplace Signal} & \textbf{Cauchy Signal} \\
    \hline
    \textbf{Gaussian Prior} & \cellcolor{blue!25}Linear (DeGroot) & \cellcolor{green!25}Saturating (Bounded Shift) & \cellcolor{green!10}Tail-Rejecting (Soft BC) \\
    \hline
    \textbf{Laplace Prior} & \cellcolor{red!25}Signal Tracking 
    & \multicolumn{1}{|@{}c@{}|}{\begin{tabular}[c]{>{\centering\arraybackslash}p{\taxonomycellwidth}}\rowcolor{green!25}Saturating ($\sigma_0 < \sigma_\epsilon$)\\
    \rowcolor{blue!25}Linear ($\sigma_0 = \sigma_\epsilon$)\\
    \rowcolor{red!25}Signal Tracking ($\sigma_0 > \sigma_\epsilon$)\end{tabular}}
    & \cellcolor{green!10}Tail-Rejecting (Soft BC) \\
    \hline
    \textbf{Cauchy Prior} & \cellcolor{red!25}Signal Tracking & \cellcolor{red!25}Signal Tracking & \cellcolor{blue!25}Linear (DeGroot) \\
    \hline
    \end{tabular}
    \caption{Principal response classes generated by different choices of prior and perceived-signal distributions, distinguished by their extreme-signal behavior. BC denotes bounded confidence.}
    \label{table:models}
\end{table}

The rest of the paper is organized as follows. Section~\ref{sec:lit} reviews the psychological basis of opinion formation, classic opinion dynamics models, and Bayesian approaches to opinion updating. Section~\ref{sec:framework} introduces the framework and derives the local-linearity and likelihood-score results. Section~\ref{sec:response_classes} develops the four response classes in the order linear updating, saturation, tail rejection, and signal tracking. Section~\ref{sec:localized_mechanisms} presents local overreaction, source-attribution attenuation, and directional reversal. Section~\ref{sec:conclusion} discusses the results and concludes. The Appendix contains detailed calculations and analytical approximations in the same order as the main text.

\section{Background and Related Work}\label{sec:lit}

\subsection{Opinion and Opinion Dynamics}
Opinion dynamics models represent an individual's belief about a subject in several ways \citep{castellano_statistical_2009, des_mesnards_detecting_2022}. The belief state is commonly a scalar on a continuous interval \citep{degroot1974reaching, friedkin_social_1990, hegselmann_opinion_2002} or a discrete state \citep{holley_ergodic_1975, sznajd-weron_opinion_2000}. These states evolve as individuals interact and process information from others \citep{friedkin_social_1990, flache_models_2017}. At the population level, opinion dynamics studies how such interactions produce consensus, disagreement, clustering, and other collective patterns.

\subsection{Psychological Foundations}
Opinion dynamics models draw on findings from social psychology and cognitive science. Relevant phenomena include \textit{social influence and conformity}, in which people adjust their opinions toward those of their social contacts \citep{festinger1954theory, asch1956studies}; \textit{belief perseverance}, or resistance to change in the face of contrary information \citep{edwards1968conservatism, redlawsk2002hot}; and \textit{confirmation bias}, in which people favor information that supports their existing beliefs and may discount sources perceived as too distant from their own position \citep{nickerson1998confirmation, iyengar2009red}. Initial opinions can also act as anchors, and responsiveness may decline with opinion distance \citep{tversky_judgment_1974, yaniv_receiving_2004}. Other documented patterns include overreaction to new information \citep{tversky_judgment_1974, de_bondt_does_1985, barberis_model_1998}. In some contexts, contrary evidence may even strengthen an existing belief, a pattern commonly called the backfire effect \citep{redlawsk2002hot, nyhan_when_2010, bail2018exposure, swire-thompson_searching_2020}. These patterns are often interpreted as departures from Bayesian rationality.

\subsection{Classic Opinion Dynamics Models}
Several mathematical models capture these phenomena. The DeGroot model \citep{degroot1974reaching} represents an opinion update as a weighted average of connected individuals' opinions. The Friedkin--Johnsen model \citep{friedkin_social_1990} adds persistent attachment to initial beliefs. In bounded-confidence models \citep{deffuant_mixing_2001, hegselmann_opinion_2002}, individuals influence one another only when their opinions are sufficiently close. Other models use amplification factors to represent overreaction \citep{barberis_model_1998}, repulsive interactions to represent backfire \citep{jager_uniformity_2005}, or nonlinear response functions to introduce further psychological realism \citep{dandekar_biased_2013}. Each formulation is useful for its intended phenomenon, but their different mechanisms make systematic comparison difficult. Table~\ref{table:model_psychology_mapping} summarizes the models and their common behavioral interpretations.

\subsection{Bayesian Opinion Updating}
Many opinion dynamics models offer valuable insights without providing a common probabilistic explanation for their update rules. Bayesian updating provides such a basis: an individual combines a prior belief with a subjective likelihood that describes the perceived signal-generating process. This receiver-side problem is distinct from Bayesian persuasion, in which a sender strategically designs an information structure \citep{kamenica2011bayesian}; our focus is how a receiver updates under a given perceived-signal distribution. The resulting rule is coherent with the receiver's subjective probabilistic model and permits behavioral patterns to arise through structured priors and subjective likelihoods \citep{gigerenzer1995improve, griffiths_optimal_2006}.

Earlier work has shown that particular opinion-update functions can arise from Bayesian specifications \citep{martins_bayesian_2009, acemoglu2011bayesian, jadbabaie_non-bayesian_2012}. Martins provides a closely related example in which a Gaussian prior and a mixture signal with an unbiased Gaussian component and a uniform state-independent component generate responses similar to bounded-confidence updating \citep{martins_bayesian_2009}. A finance application likewise combines a Gaussian prior with a power-law signal to obtain a bounded-confidence-type forecast update \citep{de_silva_expectations_2025}. Building on the general Bayesian insight, our contribution is a systematic distributional analysis across Gaussian, Laplace, Cauchy, biased, and mixture specifications. This analysis yields exact and asymptotic mappings to four principal response classes, together with local overreaction, localized attenuation, and directional reversal.

Together, these results organize seemingly disparate opinion-update functions through their prior and perceived-signal distributions. They also clarify the connections among established response patterns, provide tractable analytical approximations, and identify response forms generated by new distributional combinations.

\section{Bayesian Framework and Local Behavior}\label{sec:framework}
Opinion dynamics generally describe how opinions evolve through repeated signals and social interactions over time. This paper focuses on a single-receiver, one-time update: how an opinion changes after one observed signal. Repeated application of the derived update rules can generate temporal or network-level opinion dynamics. We leave those repeated dynamics for future work and include one sequential extension in Subsection~\ref{sec:kalman}, where a Kalman filter describes repeated signals about a time-varying state.

Consider a sender who communicates a signal realization \(x\) about an unknown state to a receiver. The state might represent a political position, an asset value, or another quantity of interest. The receiver's one-time opinion shift is
\[
\theta_1 - \theta_0 = h(x - \theta_0),
\]
where \( \theta_0 \) represents the receiver's opinion before the signal, \( \theta_1 \) is the updated opinion, and \( h(\cdot) \) is the \textit{opinion shift function}. The shift function can be linear, as in the DeGroot and Friedkin--Johnsen update rules, or nonlinear.

To give these response functions Bayesian microfoundations, we interpret \( \theta_0 \) as the receiver's prior opinion and \( \theta_1 \) as the posterior opinion after observing \( x \). Bayes' theorem determines \( \theta_1 \), allowing us to derive \( h(\cdot) \) from the receiver's prior and subjective signal model.

In the framework, the receiver begins with a \textit{prior belief} about the true state. This prior is represented by a random variable \( \Theta \) with cumulative distribution \( F_\Theta(\theta) = \mathbb{P}(\Theta \leq \theta) \). The initial opinion \(\theta_0\) is the center, or location parameter, of the prior, and \(\sigma_0\) is its scale. For the centered Gaussian and Laplace priors considered below, \(\theta_0\) is also the prior mean, median, and mode. For a Cauchy prior, the mean does not exist, so \(\theta_0\) denotes its location, median, and mode. The posterior mean exists for the Cauchy prior--likelihood combinations studied here and defines the updated opinion \(\theta_1\). The true state, denoted by \( \theta^* \in (-\infty, \infty) \), is unknown to the receiver but lies within the same domain as \( \Theta \).

The sender communicates a signal \(x\in(-\infty,\infty)\). The receiver represents the signal-generating process as
\[
X = \Theta + b + \epsilon,
\]
where \( b \in (-\infty, \infty) \) is the receiver's perceived source bias and \( \epsilon \) is perceived signal noise with proper density \( l_\epsilon(\cdot) \). The noise has location zero and scale \( \sigma_\epsilon \), reflecting the receiver's subjective uncertainty about the signal.

After receiving \(x\), the receiver updates by Bayes' rule. The posterior mean is
\begin{align*}
\theta_1 &= \mathbb{E}[\Theta | X = x] \\
&= \frac{\int_{\theta=-\infty}^{\infty} \theta \, l_\epsilon(x - b - \theta) \, dF_\Theta(\theta)}{\int_{\theta=-\infty}^{\infty} l_\epsilon(x - b - \theta) \, dF_\Theta(\theta)},
\end{align*}
where \(dF_\Theta(\theta)=f_\Theta(\theta)\,d\theta\) when \(\Theta\) has density \(f_\Theta\). The signal likelihood $l_{X|\Theta}(x|\theta)$ is therefore $l_\epsilon(x-b-\theta)$.

We analyze two regimes:
\begin{enumerate}
    \item \textbf{Small signals near the prior opinion (\( x \approx \theta_0 \)):} The update is locally linear, with slope parametrized by the DeGroot coefficient.
    \item \textbf{Large signals far from the prior opinion (\( |x-\theta_0| \) large):} The update may remain linear, saturate, vanish, or track the signal, depending on the prior and perceived-signal distributions.
\end{enumerate}
Throughout the paper, $x_0=x-b-\theta_0$ denotes the unnormalized, bias-adjusted signal deviation; in subsections that set $b=0$, this reduces to $x_0=x-\theta_0$. A normalized deviation is denoted by $z_0$. The four response classes are summarized in Table~\ref{table:models} and developed in Section~\ref{sec:response_classes}. The shift is denoted by $h(x_0)=\theta_1-\theta_0$; local overreaction, localized attenuation, and directional reversal describe finite-region mechanisms rather than additional large-signal classes.

The next two subsections preview how the signal likelihood shapes local and large-signal behavior.

\subsection{Local Linearity}
In this subsection, we set $b=0$. With perceived bias, the same analysis applies after replacing the raw signal $x$ by the bias-adjusted signal $x-b$.

To quantify how infinitesimal changes in the observation \(x\) affect the posterior mean near the prior center \(\theta_{0}\), we define the \emph{local signal score}
\begin{align*}
s_l(\theta_0; \theta) := \frac{d}{dx}\,\log l_{\epsilon}(x-\theta)|_{x=\theta_0},
\end{align*}
which measures the sensitivity of each likelihood contribution to a small shift in \(x\).

For \(x\approx\theta_{0}\), Appendix~\ref{app:local_score} shows that, under the regularity conditions permitting differentiation under the integral and for a prior and likelihood symmetric about their centers, the posterior mean admits the first-order expansion
\begin{align*}
\theta_{1}
&\approx
\theta_{0}
+ \omega\,(x-\theta_{0}),
\end{align*}
where
\begin{align*}
\omega
&=\operatorname{Cov}_{\pi_0}\!\bigl(\Theta,s_l(\theta_0;\Theta)\bigr)
=\mathbb{E}_{\pi_{0}}\bigl[(\Theta-\theta_0)s_l(\theta_0; \Theta)\bigr],
\end{align*}
and $\pi_{0}(d\theta)\propto f_{\Theta}(\theta)\,l_{\epsilon}(\theta_{0}-\theta)\,d\theta$ is the posterior at \(x=\theta_{0}\). The score \(s_l(\theta_0; \theta)\) captures how a small change in \(x\) perturbs the likelihood at each \(\theta\), and its posterior covariance with \(\Theta\) gives the instantaneous slope. Symmetry alone does not ensure that this slope lies in $[0,1]$. The unimodal Gaussian, Laplace, and Cauchy pairs studied in Subsections~\ref{sec:DeGroot}--\ref{sec:overreaction} satisfy that bound, while Subsection~\ref{sec:local_overreaction} shows that a symmetric proper prior can produce a slope greater than one.

Thus every differentiable response studied here is locally DeGroot-like. When $\omega\in[0,1]$, the local approximation is a convex combination of the prior center \(\theta_{0}\) and the signal \(x\).

\subsection{Likelihood-Score Diagnostic under a Concentrated Prior}
To isolate how the perceived-signal likelihood shapes an update, consider the concentrated uniform prior
\begin{align*}
    f_\Theta(\theta) = \begin{cases}
        (2\delta)^{-1} & \text{if} ~~|\theta - \theta_0| \leq \delta, \\
        0 & \text{otherwise},
    \end{cases}
\end{align*}
and likelihood $l(x|\theta):=l_\epsilon(x-\theta)$. Let $l'$ and $l''$ denote derivatives with respect to $\theta$, assuming the likelihood is twice differentiable on the prior support. Appendix~\ref{app:concentrated_score} gives the second-order approximation
\begin{align*}
    \theta_1-\theta_0
    &\approx
    \frac{\delta^2}{3}
    \frac{l'(x|\theta_0)}
    {l(x|\theta_0)+\frac{\delta^2}{6}l''(x|\theta_0)}.
\end{align*}
This resembles Tweedie's formula \citep{efron_tweedies_2011}, but here it is a narrow-prior approximation.

Define the \textit{signal score}
\begin{align*}
    s(\theta_0;x)
    :=\frac{l'(x|\theta_0)}{l(x|\theta_0)}.
\end{align*}
When
\begin{align*}
    \frac{\delta^2}{6}
    \left|
    \frac{l''(x|\theta_0)}{l(x|\theta_0)}
    \right|\ll1,
\end{align*}
the denominator correction is negligible and
\begin{align*}
    \theta_1-\theta_0
    &\approx \frac{\delta^2}{3}s(\theta_0;x).
\end{align*}
Table~\ref{table:scores} compares the resulting score shapes. The Gaussian score is linear, the Laplace score has constant magnitude, and the Cauchy score decays to zero. For the Laplace likelihood, the displayed derivative is understood for $x\neq\theta_0$; at $x=\theta_0$, the likelihood is not differentiable in $\theta$, while symmetry gives zero posterior shift. These shapes provide a concentrated-prior diagnostic for linear, saturating, and tail-rejecting responses. The exact large-signal class also depends on the prior tails and is established by the distribution-specific results in Section~\ref{sec:response_classes}.

For a Gaussian likelihood, the score remains exactly linear:
\begin{align*}
    s(\theta_0;x)=\frac{x-\theta_0}{\sigma_\epsilon^2}.
\end{align*}
However, the concentrated-prior reduction requires
\begin{align*}
    \frac{\delta^2}{6}
    \left|
    \frac{(x-\theta_0)^2}{\sigma_\epsilon^4}
    -\frac{1}{\sigma_\epsilon^2}
    \right|\ll1.
\end{align*}
Thus the score is linear even though, for fixed $\delta$, a bounded uniform prior eventually forces the posterior mean toward a boundary of its support. This does not affect the exact Gaussian--Gaussian linear update in Theorem~\ref{thm:gaussian_degroot}.

This preliminary analysis previews three of the four response classes. Section~\ref{sec:response_classes} derives precise shift equations for linear (DeGroot), saturating (bounded shift), tail-rejecting (soft bounded confidence), and signal-tracking updates. Section~\ref{sec:localized_mechanisms} then develops local overreaction, localized attenuation, and directional reversal.

Before turning to the individual models, note that the Gaussian, Laplace, and Cauchy priors are symmetric about $\theta_0$, while their corresponding noise densities are symmetric about zero. For unbiased, non-mixture signals, these symmetries make the shift function $h(x-\theta_0)$ odd. It therefore suffices to derive the response for $x-\theta_0\geq0$ and obtain the response for $x-\theta_0\leq0$ by odd symmetry.

\begin{table}[htbp]
    \centering
    \caption{Signal-score functions for different perceived-signal likelihoods. These scores are used in the concentrated-prior approximation and are not, by themselves, exact large-signal posterior asymptotics.}
    \label{table:scores}
    \begin{tabular}{cc}
        \toprule
        \textbf{Signal} & \textbf{Signal Score} \\
        \midrule
        Gaussian & $\displaystyle \frac{x - \theta_0}{\sigma_\epsilon^2}$ \\
        Laplace & $\displaystyle \frac{\operatorname{sgn}(x - \theta_0)}{\sigma_\epsilon}$ \\
        Cauchy & $\displaystyle \frac{2(x - \theta_0)}{(x - \theta_0)^2 + \sigma_\epsilon^2}$ \\
        \bottomrule
    \end{tabular}
\end{table}

\section{Four Response Classes}\label{sec:response_classes}

Table~\ref{table:models} and Figure~\ref{fig:shift_grid_equal} summarize the distributional taxonomy. The four classes are distinguished by their extreme-signal behavior. With $h(x_0)=\theta_1-\theta_0$, their defining properties are
\begin{align*}
    \text{linear updating:}\quad
    &h(x_0)=\omega x_0 \quad \text{for all }x_0,\\
    \text{saturation:}\quad
    &\lim_{x_0\to\pm\infty}h(x_0)=\pm c,
    \qquad 0<c<\infty,\\
    \text{tail rejection:}\quad
    &\lim_{|x_0|\to\infty}h(x_0)=0,\\
    \text{signal tracking:}\quad
    &\lim_{|x_0|\to\infty}\frac{h(x_0)}{x_0}=1.
\end{align*}
For the distributions studied here, the prior and perceived-signal distributions jointly determine the response, while their relative tails are especially informative for extreme signals. We present the classes in the order linear updating, saturation, tail rejection, and signal tracking.

\subsection{Linear Updating (DeGroot)}\label{sec:DeGroot}
The DeGroot model uses the linear update
\begin{align*}
    \theta_1 - \theta_0 &= \omega (x - b - \theta_0),
\end{align*}
where $\omega \in [0, \infty)$ is the DeGroot coefficient, which measures the strength of the response. The classic DeGroot model sets $b=0$; we allow the receiver to perceive a biased signal. In network models, repeated linear averaging can lead to consensus.

The DeGroot model emerges under specific conditions for the prior and perceived-signal distributions. When both follow Gaussian or Cauchy distributions, the resulting update is linear. Equal-scale Laplace prior and noise distributions also yield the DeGroot rule. Unequal Laplace scales instead produce either saturation or signal tracking.

Table~\ref{table:DeGroot_omega} summarizes the DeGroot coefficient $\omega$ for each distribution pair. The corresponding posterior means are derived below, with details in the Appendix.

\begin{table}[h]
\centering
\begin{tabular}{cc}
\toprule
\textbf{Prior and Signal Distributions} & \textbf{$\omega$} \\
\midrule
Both Gaussian & $\frac{\sigma_0^2}{\sigma_0^2 + \sigma_\epsilon^2}$ \\
Both Laplace (equal scales) & $\frac{1}{2}$ \\
Both Cauchy & $\frac{\sigma_0}{\sigma_0 + \sigma_\epsilon}$ \\
\bottomrule
\end{tabular}
\caption{DeGroot coefficient $\omega$ for different prior and signal distributions. For Laplace distributions, the Bayesian update follows the DeGroot rule only when the prior and noise scales are equal.}
\label{table:DeGroot_omega}
\end{table}

\subsubsection{Gaussian Prior and Signal}
The prior density and signal noise likelihood are
\begin{align*}
    f_\Theta(\theta) &= \frac{1}{\sigma_0\sqrt{2\pi}}
                \exp\left(-\frac{(\theta-\theta_0)^2}{2\sigma_0^2}\right), \\
    l_\epsilon(x-b-\theta) &= \frac{1}{\sigma_\epsilon\sqrt{2\pi}}
                \exp\left(-\frac{(x-b-\theta)^2}{2\sigma_\epsilon^2}\right).
\end{align*}

The Gaussian prior is conjugate to the Gaussian likelihood, giving
\[
\theta_1 = \left(\frac{\theta_0}{\sigma_0^2} + \frac{x-b}{\sigma_\epsilon^2}\right) 
           \left(\frac{1}{\sigma_0^2} + \frac{1}{\sigma_\epsilon^2}\right)^{-1}.
\]
The posterior mean is therefore a precision-weighted average of the prior mean and the bias-adjusted signal.

\begin{theorem}[Gaussian Prior and Signal]\label{thm:gaussian_degroot}
Let the prior belief $\Theta \sim N(\theta_0, \sigma_0^2)$ and the signal noise $\epsilon \sim N(0, \sigma_\epsilon^2)$. The opinion shift from the prior mean $\theta_0$ to the posterior mean $\theta_1$ is
\begin{align*}
    \theta_1 - \theta_0 = \omega (x - b - \theta_0),
\end{align*}
where $\omega = \frac{\sigma_0^2}{\sigma_0^2 + \sigma_\epsilon^2}$ is the DeGroot coefficient.
\end{theorem}

\subsubsection{Laplace Prior and Signal with Equal Scales}
The prior density and signal noise likelihood are
\begin{align*}
    f_\Theta(\theta) &= \frac{1}{2\sigma_0} \exp\left(-\frac{|\theta-\theta_0|}{\sigma_0}\right), \\
    l_\epsilon(x-b-\theta) &= \frac{1}{2\sigma_\epsilon} \exp\left(-\frac{|x-b-\theta|}{\sigma_\epsilon}\right).
\end{align*}

Equal scales give a linear update.

\begin{theorem}[Laplace Prior and Signal with Equal Scales]\label{thm:laplace_laplace_equal_scale}
Let the prior belief $\Theta \sim \text{Laplace}(\theta_0, \sigma_0)$ and the signal noise $\epsilon \sim \text{Laplace}(0, \sigma_\epsilon)$. If $\sigma_0 = \sigma_\epsilon$, the opinion shift from the prior mean $\theta_0$ to the posterior mean $\theta_1$ is
\begin{align*}
    \theta_1 - \theta_0 = \omega (x - b - \theta_0),
\end{align*}
where $\omega = \frac{1}{2}$ is the DeGroot coefficient.
\end{theorem}
If $\sigma_0 \neq \sigma_\epsilon$, this linear relationship no longer holds; the response instead saturates or tracks the signal.

\subsubsection{Cauchy Prior and Signal}
The prior density and signal noise likelihood are
\begin{align*}
    f_\Theta(\theta) &= \frac{1}{\pi \sigma_0 \left[1+\left(\frac{\theta-\theta_0}{\sigma_0}\right)^2\right]}, \\
    l_\epsilon(x-b-\theta) &= \frac{1}{\pi \sigma_\epsilon \left[1+\left(\frac{x-b-\theta}{\sigma_\epsilon}\right)^2\right]}.
\end{align*}

The resulting posterior mean is also linear.

\begin{theorem}[Cauchy Prior and Signal]\label{thm:cauchy_degroot}
Let the prior belief $\Theta \sim \text{Cauchy}(\theta_0, \sigma_0)$ and the signal noise $\epsilon \sim \text{Cauchy}(0, \sigma_\epsilon)$. The opinion shift from the prior location $\theta_0$ to the posterior mean $\theta_1$ is
\begin{align*}
    \theta_1 - \theta_0 = \omega (x - b -\theta_0),
\end{align*}
where $\omega = \frac{\sigma_0}{\sigma_0 + \sigma_\epsilon}$ is the DeGroot coefficient.
\end{theorem}

\subsubsection{Sequential Extension: Kalman Filter and the DeGroot Model}\label{sec:kalman}
Repeated updating about a fixed state causes posterior uncertainty to shrink. In the Gaussian model, the DeGroot coefficient therefore decays toward zero. A constant coefficient can instead arise when the latent state changes over time.

The Kalman filter formalizes this setting. A latent state $\Theta_t$ evolves over time and is estimated from repeated measurements $X_t$. Here, $\Theta_t$ is a time-varying truth and each $X_t$ is a new signal. This sequential extension illustrates one way to apply the framework beyond a single update; other repeated-signal settings remain open for future work. Consider
\begin{align*}
    \Theta_{t} & = \Theta_{t-1} + \delta_t\\
    X_t & = \Theta_t + \epsilon_t.
\end{align*}
Assume that $\delta_t\sim N(0,q)$ and $\epsilon_t\sim N(0,r)$, where $q=\sigma_\delta^2$ and $r=\sigma_\epsilon^2$. The Kalman posterior-mean update has the DeGroot form, and its time-varying gain converges to a positive constant. Appendix~\ref{app:kalman} provides the variance recursions and derivation.
\begin{theorem}[Kalman Filter Gives Fixed DeGroot Coefficient]
Assume that the true state follows the Gaussian random walk above with process-noise variance $q=\sigma_\delta^2>0$ and measurement-noise variance $r=\sigma_\epsilon^2>0$. The opinion shift from the predicted mean $\theta_{t-1}$ to the filtered mean $\theta_t$ is
\begin{align*}
    \theta_{t} - \theta_{t-1} = \omega_t (x_t - \theta_{t-1}),
\end{align*}
where the DeGroot coefficient converges to
\begin{align*}
    \omega = \frac{2}{1+\sqrt{1+4r/q}}
    =\frac{2}{1+\sqrt{1+4(\sigma_\epsilon/\sigma_\delta)^2}}.
\end{align*}
\end{theorem}

For noisy signals, where $\sigma_\epsilon \gg \sigma_\delta$, the coefficient satisfies
\begin{align*}
    \omega \sim \frac{\sigma_\delta}{\sigma_\epsilon}.
\end{align*}

In the opposite regime, where $\sigma_\delta \gg \sigma_\epsilon$,
\begin{align*}
    \omega \approx \frac{\sigma_\delta^2}{\sigma_\delta^2 + \sigma_\epsilon^2},
\end{align*}
which agrees, to this order, with the Gaussian DeGroot coefficient when the evolving-state variance plays the role of prior uncertainty. A low-noise measurement places nearly all weight on the new signal. More generally, process noise continually restores uncertainty, preventing the posterior variance and the updating coefficient from collapsing to zero. The steady-state gain therefore supplies a Bayesian basis for the constant coefficient in the DeGroot model.


\subsection{Saturation (Bounded Shift)}\label{sec:bounded_shift}
We call an update \textit{saturating} when its magnitude approaches a finite, nonzero constant for extreme signals. An idealized bounded-shift rule is
\begin{align*}
    \theta_1 - \theta_0 &= 
    \begin{cases} 
    \omega (x - \theta_0), & \text{if } |x - \theta_0| < \tau, \\
        \omega \tau\,\operatorname{sgn}(x-\theta_0), & \text{otherwise},
    \end{cases}
\end{align*}
where \(\omega\) is the local DeGroot coefficient and \(\tau\) is the threshold in signal deviation. Below the threshold, the rule is linear; beyond it, the shift is fixed in magnitude. The Bayesian response functions derived below are smooth rather than piecewise linear, but have the same defining asymptotic property.

For example, a buyer may raise their valuation as a seller's stated value increases, but only up to a limit set by comparable products, budget constraints, or prior knowledge. Negative revisions may approach a similar lower limit. In a saturating response, updating continues, but the shift remains bounded at the extremes. Bounded shifts have also been observed in sales-forecast revisions \citep{de_silva_expectations_2025}.

Saturating updates emerge when a Laplace perceived-signal distribution is paired with a Gaussian prior or with a Laplace prior of smaller scale. Even for extreme signals, the receiver continues to update, but only by a bounded amount. This differs from tail rejection, where the shift vanishes. A Laplace prior paired with Laplace noise is exactly linear only when their scales match; unequal scales instead produce either saturation or signal tracking.


\subsubsection{Laplace Prior and Signal with Unequal Scales}

The prior density and signal noise likelihood are
\begin{align*}
    f_\Theta(\theta) &= \frac{1}{2\sigma_0}
                \exp\left(-\frac{|\theta-\theta_0|}{\sigma_0}\right), \\
    l_\epsilon(x-\theta-b) &= \frac{1}{2\sigma_\epsilon}
                \exp\left(-\frac{|x-\theta-b|}{\sigma_\epsilon}\right).
\end{align*}

For convenience, set $b=0$ and define the shifted signal $x_0=x-\theta_0$, the inverse scales $a_0=1/\sigma_0$ and $a_\epsilon=1/\sigma_\epsilon$, and the offset $x^*=2a_0/(a_\epsilon^2-a_0^2)$.

\begin{theorem}[Laplace Prior, Laplace Signal with Unequal Scales]\label{thm:laplace_laplace_unequal}
Let the prior belief $\Theta \sim \text{Laplace}(\theta_0, \sigma_0)$ and the signal noise $\epsilon \sim \text{Laplace}(0, \sigma_\epsilon)$. If $\sigma_0 \neq \sigma_\epsilon$, the Bayesian posterior mean is
\[
\theta_1 = \theta_0 + \operatorname{sgn}(x_0) \frac{a_\epsilon}{a_\epsilon e^{(a_\epsilon - a_0)|x_0|} - a_0}
\left( (|x_0| - x^*) e^{(a_\epsilon - a_0)|x_0|} + x^* \right)
\]
\end{theorem}

For $x_0\geq0$, this posterior mean can be seen as the weighted sum of the shifted signal $x_0-x^*$ and the offset $x^*$; the factor $\operatorname{sgn}(x_0)$ supplies the odd extension to $x_0<0$. Which term dominates depends on the sign of $a_\epsilon-a_0$, equivalently the sign of $\sigma_0-\sigma_\epsilon$. If $\sigma_0>\sigma_\epsilon$, the posterior is dominated by the shifted signal and asymptotically tracks it. If $\sigma_0<\sigma_\epsilon$, the offset dominates and the response saturates.

To parameterize the idealized bounded-shift benchmark, we derive the local DeGroot coefficient $\omega$ and infer an effective threshold $\tau$ by dividing the limiting shift magnitude by $\omega$.

For small signals, $x \approx \theta_0$, the response is locally DeGroot.
\begin{corollary}[DeGroot Coefficient of Laplace Prior, Laplace Signal with Unequal Scales]
Let the prior belief $\Theta \sim \text{Laplace}(\theta_0, \sigma_0)$ and the signal noise $\epsilon \sim \text{Laplace}(0, \sigma_\epsilon)$. If $\sigma_0 \neq \sigma_\epsilon$, the corresponding DeGroot coefficient near $x=\theta_0$ is
\begin{align*}
    \omega 
    & = \frac{\sigma_0}{\sigma_0 + \sigma_\epsilon}.
\end{align*}
\end{corollary}
This coefficient has the same form as for a Cauchy prior and Cauchy signal noise.

For large signals, $|x_0|=|x-\theta_0| \gg \sigma_0,\sigma_\epsilon$.
\begin{corollary}[Asymptotic Shift for Laplace Prior, Laplace Signal with Unequal Scales]
Let the prior belief $\Theta \sim \text{Laplace}(\theta_0, \sigma_0)$ and the signal noise $\epsilon \sim \text{Laplace}(0, \sigma_\epsilon)$. If $\sigma_0 \neq \sigma_\epsilon$, the asymptotic shift is
\begin{align*}
    \theta_1 - \theta_0 \approx
    \begin{cases}
        x_0 - \operatorname{sgn}(x_0) x^*, & \text{if } \sigma_0 > \sigma_\epsilon, \\
        - \operatorname{sgn}(x_0)\left(\frac{\sigma_0}{\sigma_\epsilon}\right) x^*, & \text{if } \sigma_0 < \sigma_\epsilon.
    \end{cases}
\end{align*}
\end{corollary}
For $\sigma_0 > \sigma_\epsilon$, the response tracks the signal with an offset. When $\sigma_0 < \sigma_\epsilon$, the signal's greater uncertainty relative to the prior limits the opinion shift to a constant for large signals. Its magnitude is
\[
\frac{2}{1 - \left(\frac{\sigma_0}{\sigma_\epsilon}\right)^2} (\frac{\sigma_0^2}{\sigma_\epsilon}).
\]  
The limiting magnitude increases with prior uncertainty $\sigma_0$ and decreases with perceived signal uncertainty $\sigma_\epsilon$.

With Laplace priors and perceived signals, the response class depends on their relative scales. We obtain saturation when the prior scale is smaller, signal tracking when it is larger, and exact DeGroot linearity when the scales are equal.

\subsubsection{Gaussian Prior, Laplace Signal}
The prior density and signal noise likelihood are
\begin{align*}
    f_\Theta(\theta) &= \frac{1}{\sigma_0\sqrt{2\pi}}
                \exp\left(-\frac{(\theta-\theta_0)^2}{2\sigma_0^2}\right), \\
    l_\epsilon(x-\theta-b) &= \frac{1}{2\sigma_\epsilon}
                \exp\left(-\frac{|x-\theta-b|}{\sigma_\epsilon}\right).
\end{align*}

For convenience, set $b=0$ and define $a=\sqrt{2}\sigma_0/\sigma_\epsilon$ and $z_0=x_0/(\sqrt{2}\sigma_0)=(x-\theta_0)/(\sqrt{2}\sigma_0)$.

\begin{theorem}[Gaussian Prior, Laplace Signal]\label{thm:gaussian_laplace}
Let the prior belief $\Theta \sim N(\theta_0, \sigma_0^2)$ and the signal noise $\epsilon \sim \text{Laplace}(0, \sigma_\epsilon)$. The Bayesian posterior mean is
\[
\theta_1 = \theta_0 + (\frac{\sigma_0^2}{\sigma_\epsilon})\frac{e^{-az_0}\text{erfc}\paranth{\frac{a}{2}-z_0}  -
    e^{az_0}\text{erfc}\paranth{\frac{a}{2}+z_0} }
    {e^{-az_0}\text{erfc}\paranth{\frac{a}{2}-z_0}  +
    e^{az_0}\text{erfc}\paranth{\frac{a}{2}+z_0}  }
\]
\end{theorem}

The opinion shift $\theta_1-\theta_0$ is odd in $x_0$, equivalently in $z_0$. We study its local behavior near $x=\theta_0$ and its limiting behavior for large $|x-\theta_0|$. The local slope gives the DeGroot coefficient $\omega$, while the large-signal limit gives the saturation level.

For small signals, $x \approx \theta_0$, the response is locally DeGroot.
\begin{corollary}[DeGroot Coefficient of Gaussian Prior, Laplace Signal]
Let the prior belief $\Theta \sim N(\theta_0, \sigma_0^2)$ and the signal noise $\epsilon \sim \text{Laplace}(0, \sigma_\epsilon)$. Near $x=\theta_0$, the corresponding DeGroot coefficient is
\begin{align*}
    \omega = \frac{a}{2}(\frac{2}{\sqrt{\pi} \text{erfcx}(\frac{a}{2})} - a).
\end{align*}
\end{corollary}

For $a \gg 1$, equivalently $\sigma_0 \gg \sigma_\epsilon$, an inverse-factorial expansion of $\text{erfcx}$ gives
\begin{align*}
    \omega \approx 1 - \frac{4}{a^2}.
\end{align*}

For $a \ll 1$, equivalently $\sigma_0 \ll \sigma_\epsilon$, a Taylor expansion gives
\begin{align*}
    \omega \approx \frac{a}{\sqrt{\pi}}.
\end{align*}

For large signals, $|x_0|=|x-\theta_0| \gg \sigma_0,\sigma_\epsilon$, the shift approaches a constant.
\begin{corollary}[Asymptotic Shift for Gaussian Prior, Laplace Signal]
Let the prior belief $\Theta \sim N(\theta_0, \sigma_0^2)$ and the signal noise $\epsilon \sim \text{Laplace}(0, \sigma_\epsilon)$. The asymptotic shift is
\begin{align*}
    \theta_1 - \theta_0 \approx
    \operatorname{sgn}(x_0) \frac{\sigma_0^2}{\sigma_\epsilon}
\end{align*}
\end{corollary}
The shift approaches a bounded nonzero constant, placing this model in the saturating, or bounded-shift, class. The limiting magnitude is similar to that of the unequal-scale Laplace--Laplace model.


\subsection{Tail Rejection (Soft Bounded Confidence)}\label{sec:bounded_confidence}
The classical, or hard, bounded-confidence rule is
\begin{align*}
    \theta_1 - \theta_0 &= 
    \begin{cases} 
    \omega (x - \theta_0), & \text{if } |x - \theta_0| < \tau, \\
    0, & \text{otherwise},
    \end{cases}
\end{align*}
where $\omega \in [0, \infty)$ is the DeGroot coefficient and $\tau \in [0, \infty)$ is the confidence threshold. Here we let $b=0$ to simplify notation. This hard-threshold model imposes an exact zero response outside the confidence interval. The Bayesian models derived below instead generate \textit{soft bounded confidence}: the response is smooth and generally nonzero for every finite signal, but vanishes as the signal becomes extreme. We therefore classify these responses as \textit{tail-rejecting updates}.

For the distribution families studied here, tail rejection emerges when the perceived-signal distribution has heavier tails than the prior distribution. Specifically, a Gaussian or Laplace prior paired with a Cauchy perceived-signal distribution produces soft bounded confidence. Mixture signals create a related but distinct pattern of localized attenuation, studied separately below.

\subsubsection{Gaussian Prior, Cauchy Signal}
The prior density and signal noise likelihood are
\begin{align*}
    f_\Theta(\theta) &= \frac{1}{\sigma_0\sqrt{2\pi}}
                \exp\left(-\frac{(\theta-\theta_0)^2}{2\sigma_0^2}\right), \\
    l_\epsilon(x-b-\theta) &= \frac{1}{\pi \sigma_\epsilon \left[1+\left(\frac{x-b-\theta}{\sigma_\epsilon}\right)^2\right]}.
\end{align*}

For convenience, set $b=0$ and define the normalized signal deviation $z_0=x_0/\sigma_0=(x-\theta_0)/\sigma_0$ and scale ratio $a=\sigma_\epsilon/\sigma_0$.

\begin{theorem}[Gaussian Prior, Cauchy Signal]\label{thm:gaussian_cauchy}
Let the prior belief $\Theta \sim N(\theta_0, \sigma_0^2)$ and the signal noise $\epsilon \sim \text{Cauchy}(0, \sigma_\epsilon)$. The Bayesian posterior mean is
\begin{align}
    \theta_1 & = x + \sigma_\epsilon \frac{\text{Im}\left\{\text{erfcx}\left(z\right)\right\}}{\text{Re}\left\{\text{erfcx}\left(z\right)\right\}}.\label{eq:bc_gaussian}
\end{align}
where $z = \frac{a+iz_0}{\sqrt{2}}$ and $\text{erfcx}(\cdot)$ is the scaled complementary error function given by
\begin{align*}
\text{erfcx}(z) = e^{z^2} \left(1-\frac{2}{\sqrt{\pi}} \int_0^z e^{-t^2} \, dt\right).
\end{align*}
\end{theorem}

The tail-rejecting shift $\theta_1-\theta_0$ is not transparent in equation~\eqref{eq:bc_gaussian}. Near $\theta_0$, it is locally linear with a DeGroot coefficient; far from $\theta_0$, it decays inversely with the signal deviation.

Near $x=\theta_0$, the soft-bounded-confidence response is locally DeGroot.
\begin{corollary}[DeGroot Coefficient of Gaussian Prior, Cauchy Signal]
Let the prior belief $\Theta \sim N(\theta_0, \sigma_0^2)$ and the signal noise $\epsilon \sim \text{Cauchy}(0, \sigma_\epsilon)$. When $x \approx \theta_0$, the corresponding DeGroot coefficient is
\begin{align*}
    \omega 
    & = 1 +  a^2-\sqrt{\frac{2}{\pi} }\frac{a}{\text{erfcx}(a/\sqrt{2})}.
\end{align*}
\end{corollary}

Table~\ref{table:bc_degroot} also gives approximations when the perceived-signal scale is much greater or smaller than the prior scale.

\begin{table}[h!]
\centering
\begin{tabular}{|l|l|}
\toprule
\textbf{Model Regime} & \textbf{DeGroot Coefficient} \\
\hline
Gaussian prior & $1+a^2 -\sqrt{\frac{2}{\pi}}a\left(\text{erfcx}(a/\sqrt{2})\right)^{-1} $ \\\hline
Gaussian prior, $\sigma_0 \ll \sigma_\epsilon$ &$2a^{-2}$ \\\hline
Gaussian prior, $\sigma_0 \gg \sigma_\epsilon$ &  $1-\sqrt{\frac{2}{\pi}}a$ \\\hline
\bottomrule
\end{tabular}
\caption{Local DeGroot coefficient $\omega$ for the Gaussian-prior, Cauchy-signal soft-bounded-confidence response. Here $a=\sigma_\epsilon/\sigma_0$.}
\label{table:bc_degroot}
\end{table}

For large signals, $|x_0|=|x-\theta_0| \gg \sigma_0, \sigma_\epsilon$.
\begin{corollary}[Asymptotic Shift from Gaussian Prior, Cauchy Signal]
Let the prior belief $\Theta \sim N(\theta_0, \sigma_0^2)$ and the signal noise $\epsilon \sim \text{Cauchy}(0, \sigma_\epsilon)$. The asymptotic shift is
\begin{align}
    \theta_1 - \theta_0 \approx \frac{2\sigma^2_0}{x-\theta_0}.
\end{align}
\end{corollary}

For signals far from the prior mean relative to $\sigma_\epsilon$, the opinion shift approaches zero at rate $(x-\theta_0)^{-1}$, producing a tail-rejecting, soft-bounded-confidence response.

\subsubsection{Laplace Prior, Cauchy Signal}
The prior density and signal noise likelihood are
\begin{align*}
    f_\Theta(\theta) &= \frac{1}{2\sigma_0}
                \exp\left(-\frac{|\theta-\theta_0|}{\sigma_0}\right), \\
    l_\epsilon(x-b-\theta) &= \frac{1}{\pi \sigma_\epsilon \left[1+\left(\frac{x-b-\theta}{\sigma_\epsilon}\right)^2\right]}.
\end{align*}

For convenience, set $b=0$ and define $z_0=x_0/\sigma_0=(x-\theta_0)/\sigma_0$ and $a=\sigma_\epsilon/\sigma_0$.

\begin{theorem}[Laplace Prior, Cauchy Signal]\label{thm:laplace_cauchy}
Let the prior belief $\Theta \sim \text{Laplace}(\theta_0, \sigma_0)$ and the signal noise $\epsilon \sim \text{Cauchy}(0, \sigma_\epsilon)$. The Bayesian posterior mean is
\begin{align}
    \theta_1 & =  x +\sigma_\epsilon\frac{\text{Re}\left(2\pi i \cosh(z)- \text{Eix}(z) +\text{Eix}(-z)\right)}
    {\text{Im}\left(2\pi i \cosh(z)- \text{Eix}(z) +\text{Eix}(-z)\right)},\label{eq:bc_laplace}
\end{align}
where $z = z_0 + ia$ and $\text{Eix}(\cdot)$ is the scaled exponential integral
\begin{align*}
\text{Eix}(z) = e^{-z}\operatorname{Ei}(z),
\end{align*}
with $\operatorname{Ei}(\cdot)$ taken on its principal branch.
\end{theorem}

Near $x=\theta_0$, the model is locally DeGroot with the following coefficient.
\begin{corollary}[DeGroot Coefficient of Laplace Prior, Cauchy Signal]
Let the prior belief \( \Theta \sim \text{Laplace}(\theta_0, \sigma_0) \) and the signal noise \(\epsilon \sim \text{Cauchy}(0, \sigma_\epsilon) \). When \(x \approx \theta_0 \), the corresponding DeGroot coefficient is
\begin{align*}
    \omega = 1 - a \cdot
    \frac{\frac{\pi}{2} \sin(a) - \cos(a) \text{Ci}(a) - \sin(a) \text{Si}(a)}
         {\frac{\pi}{2} \cos(a) - \cos(a) \text{Si}(a) + \sin(a) \text{Ci}(a)},
\end{align*}
where \( \text{Si}(\cdot) \) and \( \text{Ci}(\cdot) \) are the sine and cosine integrals
\begin{align*}
    \text{Si}(a) & = \int_{0}^a \frac{\sin(t)}{t} \, dt, \\
    \text{Ci}(a) & = \gamma + \log(a) + \int_{0}^a \frac{\cos(t)-1}{t} \, dt,
\end{align*}
and \( \gamma \approx 0.57721 \) is Euler's constant.
\end{corollary}

For \( a \gg 1 \), corresponding to highly noisy signals with \( \sigma_\epsilon \gg \sigma_0 \), the DeGroot coefficient is approximately
\begin{align*}
    \omega \approx \frac{4}{a^2}.
\end{align*}

For \( a \ll 1 \), corresponding to low-noise signals with \( \sigma_\epsilon \ll \sigma_0 \), the coefficient is approximately
\begin{align*}
    \omega \approx 1 + \frac{2}{\pi} a \log a.
\end{align*}

For large signals, $|x-\theta_0| \gg \sigma_0,\sigma_\epsilon$, the posterior mean has the following asymptotic form.

\begin{corollary}[Asymptotic Shift for Laplace Prior, Cauchy Signal]
Let the prior belief \( \Theta \sim \text{Laplace}(\theta_0, \sigma_0) \) and the signal noise \(\epsilon \sim \text{Cauchy}(0, \sigma_\epsilon) \). The asymptotic shift is
\begin{align*}
    \theta_1 - \theta_0 \sim \frac{4\sigma_0^2}{x-\theta_0}.
\end{align*}
\end{corollary}

More generally, for a centered symmetric prior with sufficient moments and tails permitting term-by-term integration, the leading coefficient is twice the prior variance. Thus this result has the same variance-controlled asymptotic form as the Gaussian-prior case and demonstrates that, for large signals, the Cauchy tail dominates the posterior mean and produces tail rejection.

\subsection{Signal Tracking}\label{sec:overreaction}

For the distribution families studied here, signal tracking occurs when the prior distribution has heavier tails than the perceived-signal distribution. In this class, the posterior response has asymptotic unit slope:
\begin{align*}
    \frac{\theta_1-\theta_0}{x-\theta_0}\to1.
\end{align*}
Thus the posterior mean tracks the signal in relative terms, possibly with a fixed asymptotic offset. Signal tracking arises in four cases:
\begin{enumerate}
    \item a Laplace prior and Gaussian signal,
    \item a Laplace prior and signal when the prior scale exceeds the signal scale,
    \item a Cauchy prior and Gaussian signal, and
    \item a Cauchy prior and Laplace signal.
\end{enumerate}

We use the neutral term \textit{signal tracking} because it describes an asymptotic unit response without implying that the posterior overshoots the signal or exceeds a specified rational benchmark.

To compute the opinion shift in signal-tracking models, we interchange the roles of the symmetric prior and signal location families from models already studied. Denote the earlier posterior mean by $\theta_1^0(\theta_0,x,\sigma_0,\sigma_\epsilon)$. Commutativity of the two density factors gives the new posterior mean
\begin{align}
    \theta_1 = \theta_1^0(x,\theta_0,\sigma_\epsilon,\sigma_0). \label{eq:swap}
\end{align}
Equivalently, if the earlier posterior mean is written as
\begin{align*}
    \theta_1^0
    =\theta_0+h^0(x-\theta_0;\sigma_0,\sigma_\epsilon),
\end{align*}
then odd symmetry of $h^0$ gives
\begin{align*}
    \theta_1
    =x-h^0(x-\theta_0;\sigma_\epsilon,\sigma_0).
\end{align*}
Thus the local coefficient of the swapped model is $\omega=1-\omega^0$, where $\omega^0$ is the local coefficient of the original model evaluated with the scales interchanged.

\subsubsection{Laplace Prior, Gaussian Signal}
The exact posterior mean follows from equation~\eqref{eq:swap} applied to the Gaussian-prior, Laplace-signal formula in Theorem~\ref{thm:gaussian_laplace}. Its local coefficient is $\omega=1-\omega^0$, with the scales interchanged as above. For large signals, the corresponding asymptotic result from the saturating model gives
\begin{align*}
    \theta_1 \approx x + \operatorname{sgn}(\theta_0 - x)\frac{\sigma_\epsilon^2}{\sigma_0}.
\end{align*}
Thus, when $x$ is large, the posterior mean is nearly equal to $x$, with only a constant offset (negative when $x > \theta_0$, positive otherwise).

\subsubsection{Laplace Prior and Signal with Prior Scale Exceeding Signal Scale}
This signal-tracking model was derived alongside the saturating model for unequal-scale Laplace prior and signal distributions (see Subsection~\ref{sec:bounded_shift}). Here the prior uncertainty is larger than the perceived signal uncertainty.

\subsubsection{Cauchy Prior, Gaussian Signal}
\begin{figure}[htbp]
    \centering
    \includegraphics[scale=0.33]{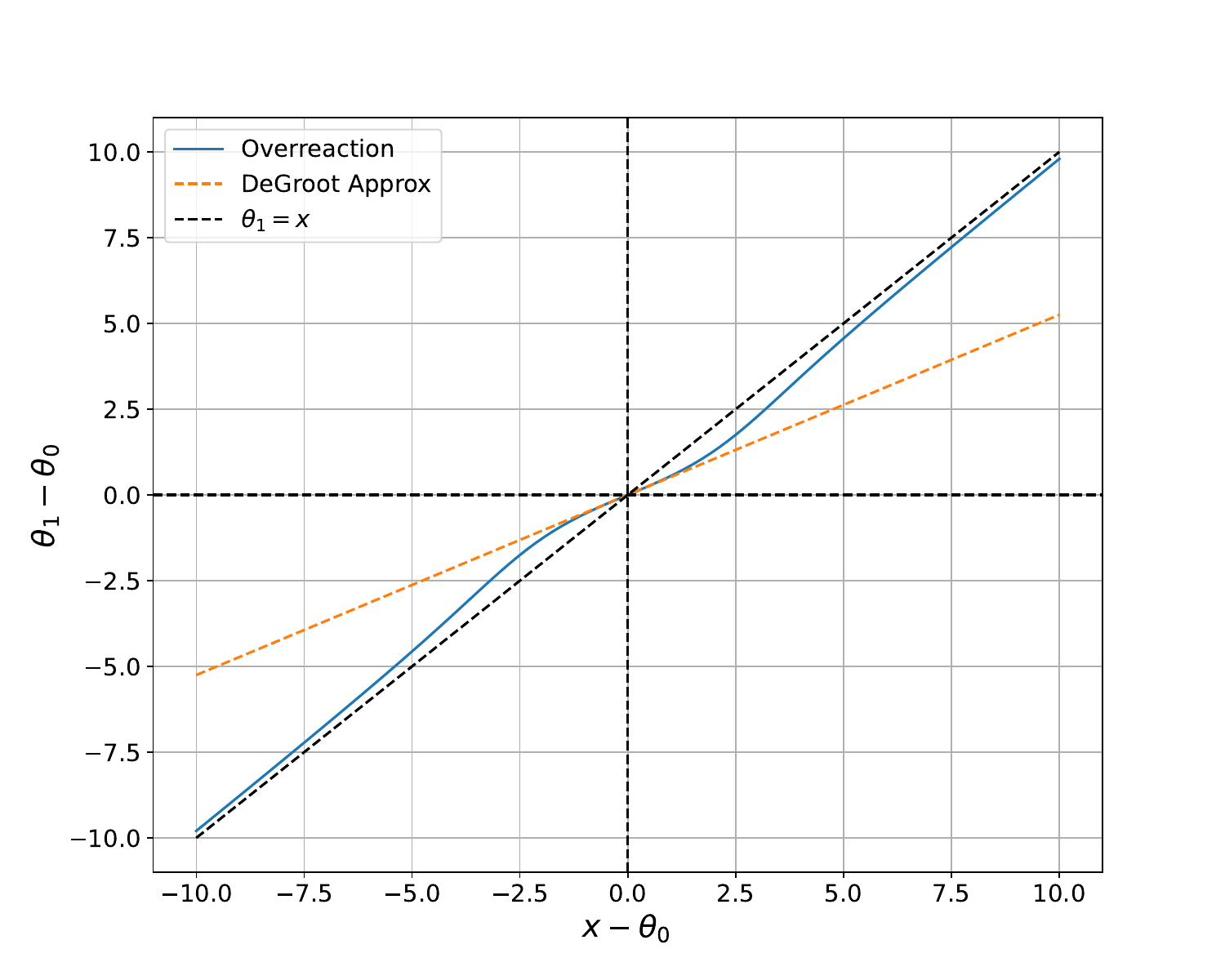}
    \caption{Opinion shift versus signal deviation for a Cauchy prior and Gaussian signal with $\sigma_0=1$ and $\sigma_\epsilon=2$. The plot includes the local DeGroot approximation and the reference line $\theta_1=x$.}
    \label{fig:signal_tracking_gaussian}
\end{figure}

Consider the tail-rejecting model with a Gaussian prior and Cauchy likelihood. Applying equation~\eqref{eq:swap} to equation~\eqref{eq:bc_gaussian} gives the posterior mean for a Cauchy prior and Gaussian signal. Figure~\ref{fig:signal_tracking_gaussian} illustrates convergence of the posterior mean to the signal for large absolute values, while small signals exhibit locally linear updating.

The exact posterior mean follows from equation~\eqref{eq:swap}. For small signals, its local coefficient is $\omega=1-\omega^0$, with $\omega^0$ evaluated for the Gaussian-prior, Cauchy-signal model after interchanging the scales. For large signals, the asymptotic behavior is
\begin{align*}
    \theta_1 \approx x + \frac{2\sigma_\epsilon^2}{\theta_0-x}.
\end{align*}
Thus the posterior mean tracks $x$ for extreme signals.

\subsubsection{Cauchy Prior, Laplace Signal}
A similar derivation applies to a Cauchy prior and Laplace signal. Its small-signal coefficient is $\omega=1-\omega^0$, with $\omega^0$ evaluated for the Laplace-prior, Cauchy-signal model after interchanging the scales. For large signals, the asymptotic behavior is
\begin{align*}
    \theta_1 \approx x + \frac{4\sigma_\epsilon^2}{\theta_0-x},
\end{align*}
so the posterior mean converges to $x$ in either tail.

\section{Localized Response Mechanisms}\label{sec:localized_mechanisms}

The next three mechanisms operate over finite regions of the response rather than define additional large-signal classes: local overreaction, source-attribution attenuation, and directional reversal.

\subsection{Local Overreaction under a Multimodal Prior}\label{sec:local_overreaction}
Consider the symmetric Gaussian-mixture prior and Gaussian perceived-signal noise
\begin{align*}
    \Theta &\sim \tfrac12 N(\theta_0-d,\sigma_0^2)
    +\tfrac12 N(\theta_0+d,\sigma_0^2),
    &\epsilon&\sim N(0,\sigma_\epsilon^2).
\end{align*}
Setting $b=0$ and hence $x_0=x-\theta_0$, Appendix~\ref{app:bimodal_overreaction} shows that the posterior mean is
\begin{align*}
    \theta_1
    =\theta_0
    +\frac{\sigma_0^2}{\sigma_0^2+\sigma_\epsilon^2}x_0
    +\frac{\sigma_\epsilon^2d}{\sigma_0^2+\sigma_\epsilon^2}
    \tanh\left(\frac{d x_0}{\sigma_0^2+\sigma_\epsilon^2}\right).
\end{align*}
At the symmetry point, $\theta_1-\theta_0=0$, but the local coefficient satisfies
\begin{align*}
    \omega_0
    =\frac{\sigma_0^2}{\sigma_0^2+\sigma_\epsilon^2}
    +\frac{d^2\sigma_\epsilon^2}{(\sigma_0^2+\sigma_\epsilon^2)^2}>1
    \quad\Longleftrightarrow\quad
    d^2>\sigma_0^2+\sigma_\epsilon^2.
\end{align*}
Thus sufficiently separated prior modes generate local overreaction around a zero-shift symmetry point. This is a localized mechanism rather than an additional large-signal class.

More generally, suppose the prior is a finite mixture $f_\Theta(\theta)=\sum_{i=1}^k\pi_i f_i(\theta)$. Appendix~\ref{app:mixture_prior} shows that its posterior mean is
\begin{align*}
    \theta_1=\sum_{i=1}^k\beta_i(x)\theta_{1,i}(x),
\end{align*}
where $\theta_{1,i}(x)$ is the posterior mean conditional on prior component $i$ and $\beta_i(x)$ is that component's posterior weight. Its response slope decomposes as
\begin{align*}
    \frac{d\theta_1}{dx}
    =\sum_{i=1}^k\beta_i(x)\omega_i(x)
    +\operatorname{Cov}_{\beta(x)}\!\left(\theta_{1,i}(x),r_i(x)\right),
\end{align*}
where $\omega_i(x)=d\theta_{1,i}(x)/dx$ and $r_i(x)$ is the score of component $i$'s marginal signal density. The covariance term captures signal-induced reweighting across prior components and can raise the total slope above one even when every within-component slope is at most one.

\subsection{Source Attribution under Mixture Signals}\label{sec:mixture_attenuation}
A distinct attenuation mechanism arises when the perceived signal is a mixture of distributions centered at the state $\Theta$ and at state-independent values $\Delta_i$. Let $f_\Theta$ denote the prior density and define the perceived-signal likelihood as
\begin{equation}\label{eq:mixture_signal}
    l_{X|\Theta}(x|\theta) = \phi_\epsilon l_\epsilon(x-\theta) + \sum_{i=1}^k \phi_i l_i(x-\Delta_i)
\end{equation}
where $0 < \phi_\epsilon \leq 1$, $0\leq\phi_i\leq1$, and $\phi_\epsilon + \sum_{i=1}^k\phi_i = 1$. The parameters $\phi_\epsilon$ and $\phi_i$ are the component weights. The condition $\phi_\epsilon>0$ ensures that the ratios below are defined.

\begin{theorem}[Mixture Signal]\label{thm:mixture_signal}
Given a prior density $f_\Theta$ with finite mean $\theta_0$ and the mixture likelihood in equation~\eqref{eq:mixture_signal}, the Bayesian posterior mean is
\[
\theta_1 = \alpha(x) \cdot \theta_1^{nm} + (1-\alpha(x)) \cdot \theta_0,
\]
where
\[
    \alpha(x) = \paranth{1 + \sum_{i=1}^k \frac{\phi_i}{\phi_\epsilon} \frac{l_i(x-\Delta_i)}{J_0(x)}}^{-1} \in [0,1],
\]
$J_0(x) = \int_{-\infty}^{\infty} f_\Theta(\theta)\, l_\epsilon(x-\theta)\, d\theta$ is the marginal likelihood of observing $x$ without mixture components, and $\theta_1^{nm}$ is the posterior mean under the non-mixture signal containing only the $\Theta$-centered component.
\end{theorem}

Thus the posterior mean under a mixture signal is a convex combination of the non-mixture posterior mean and the prior mean. The source-attribution weight $\alpha$ depends on the relative component weights $\phi_i/\phi_\epsilon$ and on the likelihoods $l_i(x-\Delta_i)$ relative to the non-mixture marginal likelihood $J_0(x)$.

If the signal $x$ is more likely to come from the state-dependent component, then $\alpha$ is larger and the posterior mean is closer to the non-mixture posterior mean. The shift is attenuated when $\alpha$ is smaller, either because a state-independent component carries greater weight $\phi_i \gg \phi_\epsilon$ or because the observation is more likely to have originated from that component. This source-attribution mechanism causes localized attenuation around a bias point; unlike tail rejection, it need not persist for extreme signals.

For example, suppose $\Theta \sim N(0, \sigma_0^2)$ and the perceived-signal likelihood is a Gaussian mixture. The component centered at $\Theta$ has variance $\sigma_\epsilon^2$, while a state-independent component centered at $\Delta_i$ has variance $\sigma_i^2$. From the Gaussian DeGroot model, $\theta_1^{nm}=\omega_{nm}x$, where $\omega_{nm}=\sigma_0^2/(\sigma_0^2+\sigma_\epsilon^2)$. The marginal likelihood $J_0$ is the convolution of the Gaussian prior and the state-dependent Gaussian likelihood, so
\begin{align*}
    J_0 & = \frac{1}{\sqrt{2\pi(\sigma_0^2 + \sigma_\epsilon^2)}} \exp\left(-\frac{x^2}{2(\sigma_0^2 + \sigma_\epsilon^2)}\right).
\end{align*}
The posterior mean under the Gaussian mixture signal is therefore
\begin{equation}\label{eq:post_Gaussian_mixture}
    \theta_1 = \omega_{nm} x
    \paranth{1 + \sum_{i=1}^k \frac{\phi_i}{\phi_\epsilon} \frac{\sqrt{\sigma_0^2 + \sigma_\epsilon^2}}{\sigma_i} \exp({\frac{x^2}{2(\sigma_0^2 + \sigma_\epsilon^2)} -\frac{(x-\Delta_i)^2}{2\sigma_i^2} })}^{-1}.
\end{equation}

Near $x=0$, the model is locally DeGroot.
\begin{corollary}[DeGroot Coefficient of Gaussian Prior, Gaussian Mixture Signal]
Let the prior belief $\Theta \sim N(0, \sigma_0^2)$ and the signal defined by~\eqref{eq:mixture_signal} be a Gaussian mixture. When \(x \approx 0 \), the corresponding DeGroot coefficient is
\begin{align*}
    \omega_m = \frac{\omega_{nm}}{1 + \sum_{i=1}^k \frac{\phi_i}{\phi_\epsilon} \frac{\sqrt{\sigma_0^2 + \sigma_\epsilon^2}}{\sigma_i} \exp(-\frac{\Delta_i^2}{2\sigma_i^2})}
\end{align*}
where $\omega_{nm} = \sigma_0^2/(\sigma_0^2+\sigma_\epsilon^2)$ is the DeGroot coefficient under a non-mixture or unbiased Gaussian signal.
\end{corollary}
The mixture coefficient is a scaled version of the non-mixture coefficient. If every state-independent component is far from the prior location relative to its scale, so that $|\Delta_i|\gg\sigma_i$ for every $i$, then $\omega_m\approx\omega_{nm}$. For small signals, the receiver assigns negligible probability to these distant components and updates approximately as under a non-mixture signal.

For large $|x|$, attenuation depends on the sign of $\sigma_i^2 - (\sigma_0^2 + \sigma_\epsilon^2)$ in equation~\eqref{eq:post_Gaussian_mixture}. Consider a Gaussian mixture consisting of a state-dependent component $N(\Theta, \sigma_\epsilon^2)$ and a state-independent component $N(\Delta_1, \sigma_1^2)$, with $\Delta_1 > 0$. If $\sigma_1^2 > \sigma_0^2 + \sigma_\epsilon^2$, the opinion shift decays exponentially for large $|x|$, producing tail rejection. If $\sigma_1^2 = \sigma_0^2 + \sigma_\epsilon^2$, the shift decays more slowly for large positive $x$, while for large negative $x$ it follows the DeGroot rule. If $\sigma_1^2 < \sigma_0^2 + \sigma_\epsilon^2$, the extreme-signal response returns to the DeGroot rule, but an intermediate region of localized attenuation remains. In the special case $\sigma_1=\sigma_\epsilon$, the exponent in equation~\eqref{eq:post_Gaussian_mixture} is positive on
\begin{align*}
    x\in\left(
    \frac{\Delta_1}{\omega_{nm}}\paranth{1-\sqrt{1-\omega_{nm}}},
    \frac{\Delta_1}{\omega_{nm}}\paranth{1+\sqrt{1-\omega_{nm}}}
    \right).
\end{align*}
This positive-exponent interval identifies the region of strongest localized attenuation. Away from its effective center $\Delta_1/\omega_{nm}$, the receiver again attributes the signal to the state-dependent component.

\begin{figure}[htbp]
    \centering
    \includegraphics[width = \textwidth]{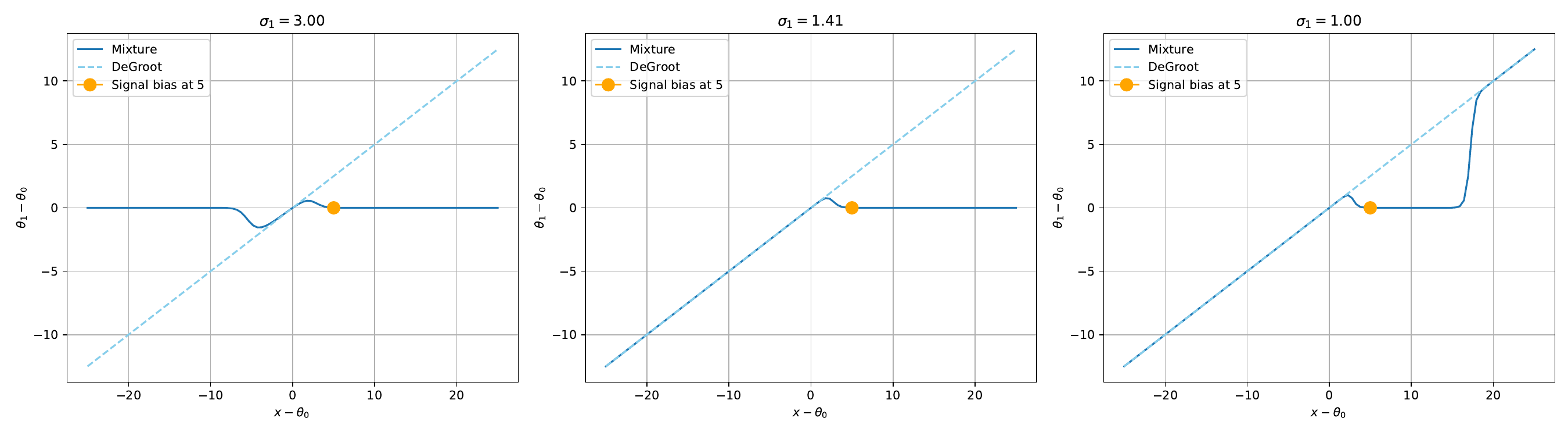}
    \caption{Opinion shift versus signal deviation for the DeGroot rule (sky blue) and the attenuated update (blue) under a Gaussian mixture signal. The state-independent component is centered at $\Delta_1=5$ (orange circle) with weight $\phi_1=0.5$. We set $\sigma_0=\sigma_\epsilon=1$. Panels show component standard deviations $\sigma_1=3$ (left), $\sqrt{2}$ (middle), and $1$ (right).}
    \label{fig:mixture_bias=5}
\end{figure}

Figure~\ref{fig:mixture_bias=5} illustrates mixture-induced localized attenuation. The prior is $N(0,1)$, and the perceived-signal likelihood is an equally weighted mixture of $N(\Theta,1)$ and $N(5,\sigma_i^2)$. Different values of $\sigma_i$ create regions in which the receiver discounts the signal by attributing it to the state-independent component. When $\sigma_i=1$, the DeGroot response returns beyond the attenuation region. Thus the mixture mechanism can be localized around the perceived bias rather than operating in both tails.

The same structure extends to uncertain state-independent centers and mixture weights. Suppose $\Delta_i \sim N(\delta_i, \tau_i^2)$ and $(\Phi_\epsilon, \Phi_1, \dots, \Phi_k) \sim \text{Dir}(\alpha_\epsilon, \alpha_1, \dots, \alpha_k)$. After integrating over these quantities, the effective likelihood remains a Gaussian mixture with adjusted means and variances. The posterior mean therefore retains the convex-combination form in Theorem~\ref{thm:mixture_signal}; the Appendix gives the derivation.


\subsection{Directional Reversal from Perceived Bias}\label{sec:backfire}
Directional reversal occurs when the sign of the opinion shift $\theta_1-\theta_0$ is opposite the sign of the raw signal deviation $x-\theta_0$. This sign pattern corresponds to a backfire effect over the region in which it occurs. Backfire effects have been reported in studies of exposure to opposing political messages \citep{bail2018exposure}, although the broader empirical literature emphasizes that such effects are context dependent \citep{swire-thompson_searching_2020}.

One way to generate directional reversal in our Bayesian framework is through a perceived fixed source bias. Consistent with Section~\ref{sec:framework}, suppose $X=\Theta+b+\epsilon$, where $b$ is the perceived bias and $\epsilon$ is zero-mean noise. First define the posterior-mean map for an unbiased signal realization $y$ as
\begin{align*}
    \mu_0(y) & =\frac{\int_{-\infty}^{\infty} \theta f_\Theta(\theta)l_\epsilon(y-\theta)d\theta}{\int_{-\infty}^{\infty}  f_\Theta(\theta)l_\epsilon(y-\theta)d\theta}.
\end{align*}
For location-family likelihoods that depend on terms of the form $x-\theta$, a biased signal is equivalent to an unbiased signal with value $x-b$. The posterior mean is therefore
\begin{align*}
    \theta_1 &  = \mu_0(x-b).
\end{align*}
Thus perceived bias translates the response function along the $x-\theta_0$ axis by $b$. This can produce directional reversal, and hence a backfire effect relative to the raw signal, over a finite region. Relative to the bias-adjusted deviation $x_0=x-b-\theta_0$, however, the receiver still moves in the direction implied by the perceived signal.

For a Gaussian prior and signal with perceived bias $b$, the opinion shift is
\begin{align*}
    \theta_1-\theta_0 & = \omega(x-b-\theta_0)=\omega x_0.
\end{align*}
For $b>0$, directional reversal occurs for $x\in(\theta_0,\theta_0+b)$: the raw signal deviation is positive but the bias-adjusted deviation, and hence the opinion shift, is negative. For $x>\theta_0+b$, the shift becomes positive and the backfire region ends. The mechanism is therefore a rational correction for perceived source bias that produces the behavioral sign pattern of backfire over a limited interval.

The magnitude of reversal depends on the perceived bias $b$ and the scales $\sigma_0$ and $\sigma_\epsilon$. Greater perceived signal uncertainty reduces the shift for a given observation, while a larger perceived bias widens the reversal region. Within the model, the partisan asymmetry reported by \citet{bail2018exposure} could reflect differences either in perceived source bias or in perceived signal noise. The first interpretation is consistent with documented partisan differences in trust in mainstream media \citep{ladd2018distrust}. These are possible model-based interpretations rather than direct empirical tests of the underlying perceptions.

\begin{figure}[htbp]
    \centering
    \includegraphics[scale = 0.33]{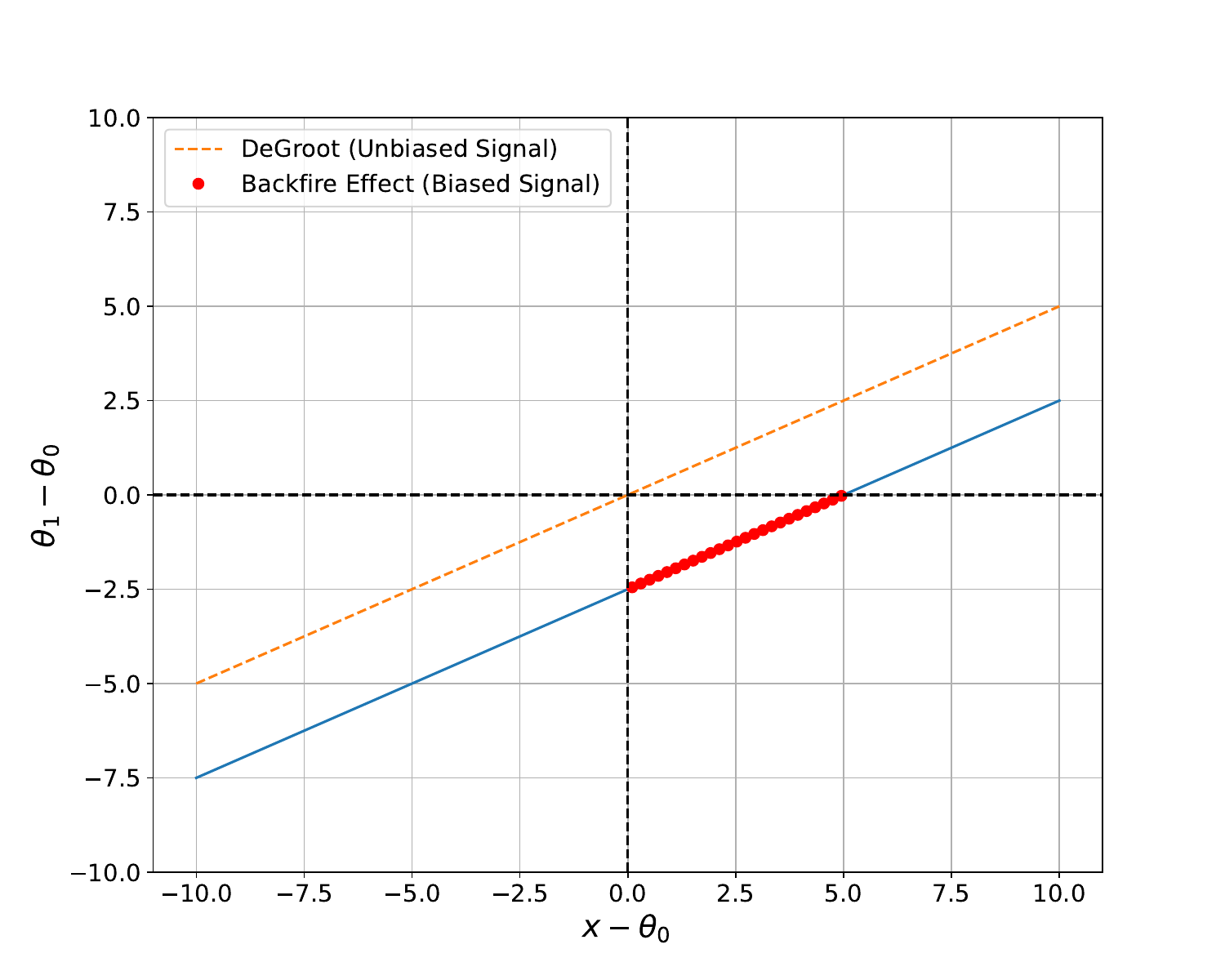}
    \caption{Opinion shift versus signal deviation for the DeGroot model with perceived bias $b=5$. All standard deviations are one, and red markers indicate the directional-reversal region.}
    \label{fig:backfire_degroot}
\end{figure}

In this model, the backfire region does not persist for sufficiently extreme signals. As Figure~\ref{fig:backfire_degroot} illustrates, the model predicts directional reversal for moderately discrepant observations but a return to movement in the signal's direction once the observation exceeds the perceived bias. This finite-region prediction is related to the ``affective tipping point'' discussed by \citet{redlawsk2010affective}.

\section{Discussion and Conclusion}\label{sec:conclusion}

We introduced a subjective Bayesian framework for the individual response functions used in opinion dynamics. The receiver interprets an observed signal through a perceived-signal distribution and updates a prior belief by Bayes' rule. Our focus is the single-receiver, one-time update. The Kalman-filter analysis provides one sequential extension; repeated signals and population-level dynamics remain important directions for future research.

Within this framework, opinion updates commonly viewed as irrational, including overreaction and bounded-confidence-type rejection, can be rational under the receiver's subjective model. Here, rationality means internal consistency with the receiver's prior, perceived signal bias, and perceived signal noise. It does not imply that this subjective model matches the objective signal-generating process.

The distributional specification determines both local and extreme-signal behavior. Near the prior opinion, the posterior-mean response is locally linear. For the distributions studied here, relative tail behavior selects among the four large-signal classes summarized in Table~\ref{table:models}. Several matched distribution pairs yield exact DeGroot rules, and the time-invariant DeGroot coefficient also arises as the steady-state gain of a Kalman filter.

Three additional mechanisms enrich this four-class taxonomy. A sufficiently separated symmetric bimodal prior produces a local response slope greater than one, and hence local overreaction, while retaining zero shift at the symmetry point. Mixture signals produce localized attenuation when the receiver attributes an observation to a state-independent component. Perceived signal bias can produce directional reversal relative to the raw signal and hence a backfire effect over a finite region determined by that bias, even though the receiver continues to move toward the bias-adjusted information. The Laplace--Laplace model is especially flexible: changing the relative scales produces saturation, exact linearity, or signal tracking.

Future work can apply these one-time response functions repeatedly over networks and characterize consensus, clustering, and polarization. Other directions include estimating the implied prior and perceived-signal distributions from behavioral data, studying asymmetric distributions, and introducing strategic signal design as a separate Bayesian-persuasion extension.

\newpage
\bibliography{Bayes}
\bibliographystyle{informs2014}

\newpage
\section{Appendix}
\subsection{Derivations for Signal Scores}
\subsubsection{Local (Small-Signal) Score}\label{app:local_score}
To characterize local, small-signal behavior near \(\theta_{0}\), define the unnormalized moments
\begin{align*}
I_{k}(x)
&=\int_{-\infty}^{\infty}
  \theta^{k}\,f_{\Theta}(\theta)\,
  l_{\epsilon}(x-\theta)\,d\theta,
  \qquad k=0,1,
\\
\theta_{1}(x)
&=\frac{I_{1}(x)}{I_{0}(x)},
\end{align*}
so that \(\theta_{1}(x)\) is the posterior mean of \(\Theta\) given the observation \(x\).

Linearizing \(\theta_{1}(x)\) about \(x=\theta_{0}\) requires the derivatives of the numerator and denominator at that point:
\begin{align*}
I_{1}'(\theta_{0})
&=\int_{-\infty}^{\infty}
  \theta\,f_{\Theta}(\theta)\,
  \left.\frac{d}{dx}l_{\epsilon}(x-\theta)\right|_{x=\theta_{0}}
  \,d\theta,
\\
I_{0}'(\theta_{0})
&=\int_{-\infty}^{\infty}
  f_{\Theta}(\theta)\,
  \left.\frac{d}{dx}l_{\epsilon}(x-\theta)\right|_{x=\theta_{0}}
  \,d\theta.
\end{align*}
These expressions measure how the likelihood moments change for an infinitesimal shift in \(x\).

The \emph{local signal score} captures the relative sensitivity of the likelihood to changes in \(x\):
\begin{align*}
s_l(\theta_0;\theta)
:&=
\left.\frac{d}{dx}\log l_{\epsilon}(x-\theta)\right|_{x=\theta_{0}}\\
&=
\frac{l_{\epsilon}'(\theta_{0}-\theta)}{l_{\epsilon}(\theta_{0}-\theta)},
\end{align*}
where the prime denotes differentiation with respect to the scalar argument of $l_\epsilon$.
By construction, this yields
\begin{align*}
I_{1}'(\theta_{0})
&=\int_{-\infty}^\infty
  \theta\,f_{\Theta}(\theta)\,
  l_{\epsilon}(\theta_{0}-\theta)\,
  s_l(\theta_0;\theta)\,d\theta,
\\
I_{0}'(\theta_{0})
&=\int_{-\infty}^\infty
      f_{\Theta}(\theta)\,
  l_{\epsilon}(\theta_{0}-\theta)\,
  s_l(\theta_0;\theta)\,d\theta.
\end{align*}

Applying the quotient rule to \(\theta_{1}(x)=I_{1}(x)/I_{0}(x)\) and evaluating at \(x=\theta_{0}\) gives
\begin{align*}
\theta_{1}'(\theta_{0})
&=\frac{I_{1}'(\theta_{0})\,I_{0}(\theta_{0})
       -I_{1}(\theta_{0})\,I_{0}'(\theta_{0})}
      {I_{0}(\theta_{0})^{2}}\\
      &= \frac{I_1'(\theta_{0})}{I_0(\theta_{0})} - \theta_1(\theta_{0})\frac{I_0'(\theta_{0})}{I_0(\theta_{0})}\\
      &=\operatorname{Cov}_{\pi_0}\!\bigl(\Theta,s_l(\theta_0;\Theta)\bigr),
\end{align*}
where \(\pi_{0}(d\theta)\propto f_{\Theta}(\theta)l_{\epsilon}(\theta_{0}-\theta)d\theta\) is the posterior at \(x=\theta_0\). If the prior is symmetric about \(\theta_0\) and the likelihood is symmetric about zero, then \(\pi_0\) is symmetric about \(\theta_0\), \(\theta_1(\theta_0)=\theta_0\), and the score \(s_l(\theta_0;\theta)\) is odd about \(\theta_0\). Consequently, \(\mathbb E_{\pi_0}[s_l(\theta_0;\Theta)]=0\), and
\begin{align*}
\theta_{1}'(\theta_{0})
&=\mathbb{E}_{\pi_{0}}\bigl[(\Theta-\theta_0)s_l(\theta_0;\Theta)\bigr].
\end{align*}

For \(x\to\theta_{0}\), the first-order Taylor expansion is
\begin{align*}
    \theta_{1}(x) &=
    \theta_{1}(\theta_{0})
    + \theta_1'(\theta_0)\,(x-\theta_{0})
    +o(|x-\theta_0|)\\
    &= \theta_0 + \omega\,(x-\theta_{0})+o(|x-\theta_0|),\\
    \omega &= \theta_{1}'(\theta_{0})
    =\mathbb{E}_{\pi_{0}}\bigl[(\Theta-\theta_0)s_l(\theta_0;\Theta)\bigr].
\end{align*}
This identity establishes local linearity but does not, from symmetry alone, imply $0\leq\omega\leq1$. The explicit Gaussian, Laplace, and Cauchy combinations in the main text have coefficients in this interval.

\subsubsection{Concentrated-Prior Signal-Score Approximation}\label{app:concentrated_score}
Suppose the uniform prior for $\Theta$ is supported on the narrow interval $[\theta_0 - \delta, \theta_0 + \delta]$. For fixed $x$ and small $\delta$, assume that the likelihood is twice differentiable on this interval. For a Laplace likelihood, this requires $\delta<|x-\theta_0|$; the case $x=\theta_0$ instead has zero shift by symmetry. The likelihood then admits a second-order Taylor approximation in $\theta$ about $\theta_0$:
\begin{align*}
    l(x|\theta) & \approx l(x|\theta_0) + \frac{d l}{d \theta}|_{\theta = \theta_0}(\theta-\theta_0) + \frac{1}{2}\frac{d^2 l}{d \theta^2}|_{\theta = \theta_0}(\theta-\theta_0)^2.
\end{align*}
The resulting posterior mean is
\begin{align*}
    \theta_1 = \frac{I_1}{I_0}
\end{align*}
where
\begin{align*}
    I_0 &\approx \frac{1}{2\delta}\int_{\theta_0 - \delta}^{\theta_0+\delta} \paranth{l(x|\theta_0) + \frac{dl}{d\theta}|_{\theta = \theta_0}(\theta-\theta_0) + \frac{1}{2}\frac{d^2l}{d\theta^2}|_{\theta = \theta_0}(\theta-\theta_0)^2} d\theta\\
        &= l(x|\theta_0) + \frac{\delta^2}{6}\frac{d^2l}{d\theta^2}|_{\theta = \theta_0},
\end{align*}
and
\begin{align*}
    I_1 &\approx \frac{1}{2\delta}\int_{\theta_0-\delta}^{\theta_0+\delta}\theta \paranth{l(x|\theta_0) + \frac{dl}{d\theta}|_{\theta = \theta_0}(\theta-\theta_0) + \frac{1}{2}\frac{d^2l}{d\theta^2}|_{\theta = \theta_0}(\theta-\theta_0)^2} d\theta\\
        &\approx \theta_0I_0 +  \frac{\delta^2}{3}\frac{dl}{d\theta}|_{\theta = \theta_0}.
\end{align*}
Substituting these expressions gives
\begin{align*}
    \theta_1 &\approx \frac{\theta_0I_0 +  \frac{\delta^2}{3}l'|_{\theta=\theta_0}}{I_0}\\
    &= \theta_0 + \frac{\delta^2}{3} \frac{l'|_{\theta=\theta_0}}{l|_{\theta=\theta_0} + \frac{\delta^2}{6}l''|_{\theta=\theta_0}}\\
    &\approx \theta_0 + \frac{\delta^2}{3} \frac{l'|_{\theta=\theta_0}}{l|_{\theta=\theta_0}}\\
    &= \theta_0 + \frac{\delta^2}{3}\frac{d}{d\theta}\log(l)|_{\theta=\theta_0},
\end{align*}
where the second approximation requires
\begin{align*}
    \frac{\delta^2}{6}
    \left|
    \frac{l''(x|\theta_0)}{l(x|\theta_0)}
    \right|\ll1.
\end{align*}
The derivative $\frac{d}{d\theta}\log l(x|\theta)|_{\theta=\theta_0}$ is the signal score. This is a concentrated-prior ($\delta\to0$ at fixed $x$) approximation, not a large-$|x|$ asymptotic result for fixed $\delta$.

\subsection{Derivations for Linear Updating}
\subsubsection{Laplace Prior and Signal with Equal Scales}
With equal scales, let $\sigma_0 = \sigma_\epsilon = \sigma$. The relevant integrals are
\begin{align*}
    I_0 & = \int_{-\infty}^{\infty}e^{-\frac{|\theta-\theta_0|}{\sigma}-\frac{|x-\theta|}{\sigma}}  d\theta\\
    I_1 & = \int_{-\infty}^{\infty} \theta e^{-\frac{|\theta-\theta_0|}{\sigma}-\frac{|x-\theta|}{\sigma}}  d\theta.
\end{align*}

Define $u = \theta-\theta_0$, $x_0 = x-\theta_0$, and $a = 1/\sigma$. Then
\begin{align*}
    J_0 & = \int_{-\infty}^{\infty}e^{-a|u|-a|x_0-u|}  du\\
    J_1 & = \int_{-\infty}^{\infty}u e^{-a|u|-a|x_0-u|}  du,
\end{align*}
where $I_0 = J_0$ and $I_1 = J_1 + \theta_0 J_0$.

For $x_0\geq0$, simple integration shows
\begin{align*}
    J_0 & = \frac{e^{-ax_0}(1+ax_0)}{a}\\
    J_1 & = \frac{x_0 e^{-ax_0}(1+ax_0)}{2a}.
\end{align*}

Therefore,
\begin{align*}
    \theta_1 - \theta_0 &= \frac{I_1}{I_0} - \theta_0\\
    &= \frac{J_1}{J_0}\\
    &= \frac{x_0}{2}\\
    & = \frac{1}{2} (x-\theta_0).
\end{align*}
Odd symmetry supplies the same identity for $x_0<0$, so this is the DeGroot model with $\omega = \frac{1}{2}$.

\subsubsection{Cauchy Prior and Signal}

The relevant integrals are
\begin{align*}
    I_0 & = \int_{-\infty}^{\infty} \frac{d\theta}{((\theta-\theta_0)^2+
    \sigma_0^2)((x-\theta)^2+    \sigma_\epsilon^2)}\\
    I_1 & = \int_{-\infty}^{\infty} \frac{\theta d\theta}{((\theta-\theta_0)^2+
    \sigma_0^2)((x-\theta)^2+    \sigma_\epsilon^2)}.
\end{align*}

Define $u = \theta -\theta_0$ and $x_0 = x-\theta_0$, and let
\begin{align*}
    J_0 & = \int_{-\infty}^{\infty} \frac{du}{(u^2+
    \sigma_0^2)((u-x_0)^2+\sigma_\epsilon^2)}\\
    J_1 & = \int_{-\infty}^{\infty}  \frac{udu}{(u^2+
    \sigma_0^2)((u-x_0)^2+\sigma_\epsilon^2)}.
\end{align*}
With these simplified integrals, the original integrals are $I_0 = J_0$ and $I_1 = J_1 + \theta_0 J_0$.

Using a semicircular contour in the upper half-plane, the poles are $i\sigma_0$ and $x_0+i\sigma_\epsilon$. Direct evaluation of their residues gives
\begin{align*}
    J_0
    &=\frac{\pi(\sigma_0+\sigma_\epsilon)}
    {\sigma_0\sigma_\epsilon
    \left[x_0^2+(\sigma_0+\sigma_\epsilon)^2\right]},\\
    J_1
    &=\frac{\pi x_0}
    {\sigma_\epsilon
    \left[x_0^2+(\sigma_0+\sigma_\epsilon)^2\right]}.
\end{align*}
Consequently,
\begin{align*}
    \theta_1-\theta_0 &= \frac{I_1}{I_0} - \theta_0\\
                    & = \frac{J_1}{J_0}\\
                    & = \frac{\sigma_0 x_0}{\sigma_0+\sigma_\epsilon}\\
                    & = \frac{\sigma_0}{\sigma_0+\sigma_\epsilon}(x-\theta_0),
\end{align*}
which is the DeGroot model with $\omega = \frac{\sigma_0}{\sigma_0+\sigma_\epsilon}$.

\subsubsection{Kalman Filter}\label{app:kalman}
The Kalman filter has a latent state $\Theta_t$ that evolves over time and is estimated from repeated measurements $X_t$. In our interpretation, $\Theta_t$ is the time-varying truth and each $X_t$ is a new signal. The dynamics are
\begin{align*}
    \Theta_{t} & = \Theta_{t-1} + \delta_t\\
    X_t & = \Theta_t + \epsilon_t.
\end{align*}

Let $\delta_t\sim N(0,q)$ and $\epsilon_t\sim N(0,r)$, where $q=\sigma_\delta^2$ and $r=\sigma_\epsilon^2$. If $P_{t-1}$ is the filtered variance after observing $X_{t-1}$, then the predicted variance, Kalman gain, mean update, and filtered variance are
\begin{align*}
    P_t^- &= P_{t-1}+q,\\
    \omega_t &= \frac{P_t^-}{P_t^-+r},\\
    \theta_t &= \theta_{t-1}+\omega_t(x_t-\theta_{t-1}),\\
    P_t &= (1-\omega_t)P_t^-
    =\frac{r(P_{t-1}+q)}{P_{t-1}+q+r}.
\end{align*}
At a steady state $P_t=P_{t-1}=P$, the variance recursion gives
\begin{align*}
    P^2+qP-qr&=0,\\
    P&=\frac{q}{2}\left(\sqrt{1+4r/q}-1\right).
\end{align*}
The steady-state predicted variance is $P^-=P+q$, so the fixed gain is
\begin{align*}
    \omega
    &=\frac{P+q}{P+q+r}
    =\frac{2}{1+\sqrt{1+4r/q}}.
\end{align*}

\subsection{Derivations for Saturating Updates (Bounded Shift)}
\subsubsection{Laplace Prior and Signal with Unequal Scales}
The relevant integrals are
\begin{align*}
I_0 &= \int_{-\infty }^{\infty } \exp \paranth{-\frac{|\theta-\theta_0|}{\sigma_0}} \exp \paranth{-\frac{|x-\theta|}{\sigma_\epsilon}} \, d\theta\\
 I_1 &=  \int_{-\infty }^{\infty }\theta \exp \paranth{-\frac{|\theta-\theta_0|}{\sigma_0}} \exp \paranth{-\frac{|x-\theta|}{\sigma_\epsilon}} \, d\theta.
\end{align*}

Define $u = \theta-\theta_0$, $x_0 = x -\theta_0$, $a_0 = 1/\sigma_0$, and $a_\epsilon = 1/\sigma_\epsilon$. The simplified integrals are
\begin{align*}
    J_0 & = \int_{-\infty}^{\infty}e^{-a_0|u|-a_\epsilon|u-x_0|}  du\\
    J_1 & = \int_{-\infty}^{\infty}u e^{-a_0|u|-a_\epsilon|u-x_0|}  du
\end{align*}

For $x_0=x-\theta_0\geq0$, integrate separately over the three regions of $u$ to obtain
\begin{align*}
    J_0 & = \int_{-\infty}^{0}e^{a_0 u - a_\epsilon(x_0 - u)} du + \int_{0}^{x_0}e^{-a_0 u -a_\epsilon(x_0- u)} du + \int_{x_0}^{\infty}e^{-a_0 u -a_\epsilon(u -x_0)} du\\
    &= e^{-a_\epsilon x_0}\frac{1}{a_\epsilon+a_0} + e^{-a_\epsilon x_0}\frac{e^{(a_\epsilon-a_0)x_0} - 1}{a_\epsilon-a_0} + \frac{e^{-a_0 x_0}}{a_\epsilon+a_0}\\
    &= \frac{2}{a_\epsilon^2 - a_0^2} \left( a_\epsilon e^{-a_0 x_0} - a_0 e^{-a_\epsilon x_0} \right)\\
    \\
    J_1 & = \int_{-\infty}^{0} u e^{a_0 u - a_\epsilon(x_0 - u)} du + \int_{0}^{x_0} u e^{-a_0 u -a_\epsilon(x_0- u)} du + \int_{x_0}^{\infty} u e^{-a_0 u -a_\epsilon(u -x_0)} du\\
    &= e^{-a_\epsilon x_0}\frac{-1}{(a_\epsilon+a_0)^2} + e^{-a_\epsilon x_0}\frac{e^{(a_\epsilon-a_0)x_0}(1 - (a_\epsilon - a_0)x_0) - 1}{(a_\epsilon-a_0)^2} + \frac{e^{-a_0 x_0}(1+(a_\epsilon + a_0)x_0)}{(a_\epsilon+a_0)^2}\\
    & = \frac{2a_\epsilon}{(a_\epsilon^2-a_0^2)^2}
             \paranth{[x_0(a_\epsilon^2-a_0^2) -2a_0] e^{-a_0 x_0} + 2a_0 e^{-a_\epsilon x_0}}
\end{align*}

Therefore,
\begin{align*}
    \theta_1& = \theta_0 + \frac{J_1}{J_0}\\
            & = \theta_0 + \frac{a_\epsilon}{a_\epsilon e^{(a_\epsilon-a_0)x_0} -a_0}\paranth{(x_0-x^*)e^{(a_\epsilon-a_0)x_0}+x^*}
\end{align*}
where $x^* = \frac{2a_0}{a_\epsilon^2-a_0^2}$.

A Taylor expansion at $x_0=0$ gives
\begin{align*}
    \theta_1-\theta_0 \approx \omega(x-\theta_0),
\end{align*}
so the local DeGroot coefficient is $\omega=\sigma_0/(\sigma_0+\sigma_\epsilon)$, the same form as for the Cauchy prior and likelihood.

For large signals $x_0 \gg 0$, the behavior depends on the sign of $\sigma_0-\sigma_\epsilon$.  

If $\sigma_0 > \sigma_\epsilon$, meaning that the prior has greater uncertainty than the signal, then $a_\epsilon-a_0>0$ and
\begin{align*}
    \theta_1 &\approx \theta_0 + (x_0 - x^*)\\
    &= x - x^*.
\end{align*}
For $x \leq \theta_0$ or $x_0 \leq 0$, odd symmetry gives $\theta_1 \approx x+x^*$. This is signal tracking with a fixed offset $x^*$.

If $\sigma_0 < \sigma_\epsilon$, meaning that the prior has less uncertainty than the signal, then $a_\epsilon-a_0<0$ and
\begin{align*}
    \theta_1-\theta_0 &\approx (\frac{a_\epsilon}{-a_0}) x^*\\
    &= -(\frac{\sigma_0}{\sigma_\epsilon}) x^*.
\end{align*}
For $x \leq \theta_0$ or $x_0 \leq 0$, odd symmetry gives $\theta_1-\theta_0\approx(\sigma_0/\sigma_\epsilon)x^*$. The shift has a finite limit and is therefore saturating.

\subsubsection{Gaussian Prior and Laplace Signal}
The relevant integrals are
\begin{align*}
I_0 &= \int_{-\infty }^{\infty }\exp \paranth{-\frac{(\theta-\theta_0)^2}{2\sigma_0^2}} \exp \paranth{-\frac{|x-\theta|}{\sigma_\epsilon}} \, d\theta\\
 I_1 &=  \int_{-\infty }^{\infty }\theta\exp \paranth{-\frac{(\theta-\theta_0)^2}{2\sigma_0^2}} \exp \paranth{-\frac{|x-\theta|}{\sigma_\epsilon}} \, \, d\theta.
\end{align*}
Define $u=(\theta-\theta_0)/(\sqrt{2}\sigma_0)$, $z_0=x_0/(\sqrt{2}\sigma_0)=(x-\theta_0)/(\sqrt{2}\sigma_0)$, and $a=\sqrt{2}\sigma_0/\sigma_\epsilon$. The normalized integrals are
\begin{align*}
    J_0 & =  \int_{-\infty }^{\infty }\exp \paranth{-u^2}\exp \paranth{-a|z_0-u|} \, du\\
    J_1 & =  \int_{-\infty }^{\infty } u\exp \paranth{-u^2}\exp \paranth{-a|z_0-u|} \, du
\end{align*}
Then $I_0$ and $I_1$ reduce to
\begin{align*}
    I_0 & =  J_0\\
    I_1 & =  \theta_0J_0 + \sigma_0\sqrt{2}J_1.
\end{align*}
Direct integration gives
\begin{align*}
    J_0 & = \frac{\sqrt{\pi}e^{a^2/4}}{2}\paranth{e^{-az_0}\text{erfc}\paranth{\frac{a}{2}-z_0}  +
    e^{az_0}\text{erfc}\paranth{\frac{a}{2}+z_0}  }\\
    J_1 & = \frac{a\sqrt{\pi}e^{a^2/4}}{4}\paranth{e^{-az_0}\text{erfc}\paranth{\frac{a}{2}-z_0}  -
    e^{az_0}\text{erfc}\paranth{\frac{a}{2}+z_0}  }
\end{align*}
Taking their ratio gives
\begin{align*}
    \frac{J_1}{J_0} & = \paranth{\frac{a}{2}}\frac{e^{-az_0}\text{erfc}\paranth{\frac{a}{2}-z_0}  -
    e^{az_0}\text{erfc}\paranth{\frac{a}{2}+z_0} }
    {e^{-az_0}\text{erfc}\paranth{\frac{a}{2}-z_0}  +
    e^{az_0}\text{erfc}\paranth{\frac{a}{2}+z_0}  }
\end{align*}

Hence
\begin{align*}
    \theta_1 &= \frac{I_1}{I_0}\\
    &= \theta_0 + \sqrt{2} \sigma_0 \frac{J_1}{J_0}\\
    &= \theta_0 + (\frac{\sigma_0^2}{\sigma_\epsilon})\frac{e^{-az_0}\text{erfc}\paranth{\frac{a}{2}-z_0}  -
    e^{az_0}\text{erfc}\paranth{\frac{a}{2}+z_0} }
    {e^{-az_0}\text{erfc}\paranth{\frac{a}{2}-z_0}  +
    e^{az_0}\text{erfc}\paranth{\frac{a}{2}+z_0}  }
\end{align*}

For $x\approx\theta_0$, a Taylor expansion at $z_0=0$ gives
\begin{align*}
    \theta_1 - \theta_0 
    &= (\frac{\sigma_0^2}{\sigma_\epsilon})(\frac{2 e^{-a^2/4}}{\sqrt{\pi} \text{erfc}(\frac{a}{2})} - a)z_0 + O(z_0^3) \\
    &\approx \frac{a}{2}(\frac{2 e^{-a^2/4}}{\sqrt{\pi} \text{erfc}(\frac{a}{2})} - a)(x-\theta_0),
\end{align*}
and hence the DeGroot coefficient is
\[
\omega = \frac{a}{2}\left(\frac{2}{\sqrt{\pi} \text{erfcx}(a/2)} - a\right).
\]

For $a \gg 1$, which corresponds to $\sigma_0 \gg \sigma_\epsilon$, the standard expansion gives
\begin{align*}
    \text{erfcx}(a/2)
    =\frac{2}{\sqrt{\pi}a}
    \left(1-\frac{2}{a^2}+\frac{12}{a^4}+O(a^{-6})\right).
\end{align*}
Substitution into the exact coefficient yields
\begin{align*}
    \omega
    =1-\frac{4}{a^2}+O(a^{-4}).
\end{align*}

For $a \ll 1$, which corresponds to $\sigma_0 \ll \sigma_\epsilon$, Taylor expansion of the exact coefficient gives
\begin{align*}
    \omega=\frac{a}{\sqrt{\pi}}+O(a^2).
\end{align*}

For $x-\theta_0\to+\infty$, the part of the posterior with $\theta>x$ is asymptotically negligible. On the remaining region, the Laplace likelihood is proportional, as a function of $\theta$, to $\exp(\theta/\sigma_\epsilon)$. Exponential tilting of the $N(\theta_0,\sigma_0^2)$ prior therefore gives a limiting posterior mean $\theta_0+\sigma_0^2/\sigma_\epsilon$. Hence
\begin{align*}
    \theta_1-\theta_0
    \longrightarrow \frac{\sigma_0^2}{\sigma_\epsilon}
    \qquad\text{as }x-\theta_0\to+\infty.
\end{align*}
Odd symmetry gives the negative-tail limit $-\sigma_0^2/\sigma_\epsilon$.

This asymptotic shift has constant magnitude $\sigma_0^2/\sigma_\epsilon$, placing it in the saturating (bounded-shift) class.

\subsection{Derivations for Tail-Rejecting Updates (Soft Bounded Confidence)}
\subsubsection{Gaussian Prior, Cauchy Signal}
The posterior mean is
\begin{align*}
    \theta_1 & = \frac{I_1}{I_0}
\end{align*}
where
\begin{align*}
I_0 &= \int_{-\infty }^{\infty } \frac{\exp \left(-(\theta-\theta_0)^2/(2\sigma_0^2)\right)}{(x-\theta)^2+\sigma_\epsilon^2} \, d\theta\\
 I_1 &=  \int_{-\infty }^{\infty } \frac{\theta \exp \left(-(\theta-\theta_0)^2/(2\sigma_0^2)\right)}{(x-\theta)^2+\sigma_\epsilon^2} \, d\theta.
\end{align*}

Let $u = (\theta - \theta_0)/\sigma_0$, and define $z_0=x_0/\sigma_0=(x-\theta_0)/\sigma_0$ and $a=\sigma_\epsilon/\sigma_0$. The required normalized integrals are
\begin{align*}
J_0 &= \int_{-\infty }^{\infty } \frac{\exp\paranth{-u^2/2}}{(u-z_0)^2+a^2}du= \frac{\pi}{a}\text{Re}\curly{\exp\paranth{z^2}\text{erfc}\paranth{z}  }\\
J_1 &= \int_{-\infty }^{\infty } \frac{u\exp\paranth{-u^2/2}}{(u-z_0)^2+a^2}du= -\frac{\pi\sqrt{2}}{a}\text{Re}\curly{iz\exp\paranth{z^2}\text{erfc}\paranth{z}  }
\end{align*}
where $z = (a+iz_0)/\sqrt{2}$ and $\text{erfc}(\cdot)$ is the complementary error function given by
\begin{align*}
\text{erfc}(x) = 1-\frac{2}{\sqrt{\pi}} \int_0^x e^{-t^2} \, dt.
\end{align*}

The original integrals become
\begin{align*}
I_0 &= \frac{1}{\sigma_0}J_0\\
I_1 &= \frac{\theta_0}{\sigma_0}J_0 + J_1
\end{align*}

Substitution and simplification give
\begin{align*}
    \theta_1 & = \frac{I_1}{I_0}\\
    & = \theta_0 + \sigma_0 \frac{J_1}{J_0}\\
    & = \theta_0 + \sigma_0 \frac{-\frac{\pi\sqrt{2}}{a}\text{Re}\curly{iz\exp\paranth{z^2}\text{erfc}\paranth{z}} }{\frac{\pi}{a}\text{Re}\curly{\exp\paranth{z^2}\text{erfc}\paranth{z}  }}\\
    & = \theta_0 - \sigma_0\frac{\text{Re}\curly{(-z_0+ia)\exp\paranth{z^2}\text{erfc}\paranth{z}}}{\text{Re}\curly{\exp\paranth{z^2}\text{erfc}\paranth{z}}}\\
    & = \theta_0 + \sigma_0 z_0 + \sigma_0a \frac{\text{Im}\curly{\exp\paranth{z^2}\text{erfc}\paranth{z}}}{\text{Re}\curly{\exp\paranth{z^2}\text{erfc}\paranth{z}}}\\
    & = x + \sigma_\epsilon \frac{\text{Im}\curly{\exp\paranth{z^2}\text{erfc}\paranth{z}}}{\text{Re}\curly{\exp\paranth{z^2}\text{erfc}\paranth{z}}}\\
    & = x + \sigma_\epsilon \frac{\text{Im}\curly{\text{erfcx}\paranth{z}}}{\text{Re}\curly{\text{erfcx}\paranth{z}}},
\end{align*}
where $\text{erfcx}(z) = e^{z^2}\text{erfc}(z)$ is the scaled complementary error function. Thus, for a Gaussian prior and a Cauchy likelihood,
\begin{align*}
    \theta_1 & = x + \sigma_\epsilon \frac{\text{Im}\curly{\text{erfcx}\paranth{z}}}{\text{Re}\curly{\text{erfcx}\paranth{z}}}.
\end{align*}

For small signals, we derive an approximation for the DeGroot coefficient. We expand $\text{erfc}(\cdot)$ as
\begin{align*}
    \text{erfc}\paranth{\frac{a+iz_0}{\sqrt{2}}} = \text{erfc}\paranth{\frac{a}{\sqrt{2}}} 
    &- \frac{2}{\sqrt{\pi}}e^{-a^2/2}\int_0^{z_0/\sqrt{2}}du e^{u^2}\sin\paranth{\sqrt{2}au}\\
    &- i\frac{2}{\sqrt{\pi}}e^{-a^2/2}\int_0^{z_0/\sqrt{2}}du e^{u^2}\cos\paranth{\sqrt{2}au}
\end{align*}
using a contour from 0 to $\frac{a+iz_0}{\sqrt{2}}$.

Write $\text{erfc}(z)=A+iB$. Near $x=\theta_0$, equivalently $z_0=0$,
\begin{align*}
    A &\approx \text{erfc}(\frac{a}{\sqrt{2}}),\\
    \frac{d A}{d z_0} &= -\frac{2}{\sqrt{\pi}} e^{\frac{z_0^2 - a^2}{2}} \sin{a z_0} \approx 0,\\
    B &\approx 0,\\
    \frac{d B}{d z_0} &= -\frac{2}{\sqrt{\pi}} e^{\frac{z_0^2 - a^2}{2}} \cos{a z_0} \approx -\frac{2}{\sqrt{\pi}} e^{-\frac{a^2}{2}},\\
    \frac{d z_0}{d x} &= \frac{1}{\sigma_0}.
\end{align*}

Since the local DeGroot coefficient is $\omega=d\theta_1/dx$,
\begin{align*}
    \omega & = 1 + \sigma_\epsilon \frac{d}{d x} \frac{\text{Im}\curly{\text{erfcx}\paranth{z}}}{\text{Re}\curly{\text{erfcx}\paranth{z}}}\\
    & = 1 + \sigma_\epsilon \frac{d}{d x} \frac{\text{Im}\curly{e^{iaz_0} (A + iB)}}{{\text{Re}\curly{e^{iaz_0} (A + iB)}}}\\
    & = 1 + \frac{\sigma_\epsilon}{\sigma_0} \frac{d}{d z_0} (\frac{\sin(az_0) A + \cos(az_0) B}{\cos(az_0) A - \sin(az_0) B})\\
    & \approx 1 +  a^2-\sqrt{\frac{2}{\pi} }\frac{ae^{-a^2/2}}{\text{erfc}(a/\sqrt{2})}\\
    & = 1 +  a^2-\sqrt{\frac{2}{\pi} }\frac{a}{\text{erfcx}(a/\sqrt{2})}.
\end{align*}
The heavy-tailed Cauchy likelihood therefore gives a more involved local coefficient than the Gaussian--Gaussian model.

For $a \gg 1$, which corresponds to $\sigma_\epsilon \gg \sigma_0$, the standard large-argument expansion gives
\begin{align*}
    \text{erfcx}(a/\sqrt{2})
    =\sqrt{\frac{2}{\pi}}\frac{1}{a}
    \left(1-\frac{1}{a^2}+\frac{3}{a^4}+O(a^{-6})\right).
\end{align*}
Substitution into the exact coefficient gives
\begin{align*}
    \omega
    &=\frac{2}{a^2}+O(a^{-4})
    =\frac{2\sigma_0^2}{\sigma_\epsilon^2}
    +O\left(\frac{\sigma_0^4}{\sigma_\epsilon^4}\right).
\end{align*}

In the opposite regime where $\sigma_\epsilon \ll \sigma_0$ or $a \ll 1$, we use the Taylor expansion of $\text{erfc}(z)$ near zero:
\begin{align*}
    \text{erfc}(z) \approx 1-\frac{2}{\sqrt{\pi}} z.
\end{align*}
In this regime, expansion of the exact coefficient gives
\begin{align*}
    \omega
    &=1-\sqrt{\frac{2}{\pi}}a
    +\left(1-\frac{2}{\pi}\right)a^2+O(a^3)\\
    &=1-\sqrt{\frac{2}{\pi}}\frac{\sigma_\epsilon}{\sigma_0}
    +O\left(\frac{\sigma_\epsilon^2}{\sigma_0^2}\right).
\end{align*}

For the large-signal behavior, let $U=\Theta-\theta_0$ and $y=x-\theta_0$. Apart from a factor independent of $U$, the Cauchy likelihood is
\begin{align*}
    L_y(U)=\frac{1}{(y-U)^2+\sigma_\epsilon^2}.
\end{align*}
For large positive $y$,
\begin{align*}
    L_y(U)
    &=\frac{1}{y^2}
    \left(1-\frac{2U}{y}
    +\frac{U^2+\sigma_\epsilon^2}{y^2}\right)^{-1}\\
    &=\frac{1}{y^2}
    \left[1+\frac{2U}{y}
    +\frac{3U^2-\sigma_\epsilon^2}{y^2}
    +O(y^{-3})\right].
\end{align*}
The Gaussian prior permits term-by-term integration. Since it is centered and symmetric,
\begin{align*}
    \theta_1-\theta_0
    &=\frac{\mathbb E[U L_y(U)]}{\mathbb E[L_y(U)]}\\
    &=\frac{2\mathbb E[U^2]}{y}+O(y^{-3})\\
    &=\frac{2\sigma_0^2}{x-\theta_0}
    +O\left((x-\theta_0)^{-3}\right).
\end{align*}
Odd symmetry supplies the corresponding negative-tail expansion.

\subsubsection{Laplace Prior, Cauchy Signal}
The posterior mean is
\begin{align*}
    \theta_1 & = \frac{I_1}{I_0}
\end{align*}
where
\begin{align*}
I_0 &= \int_{-\infty }^{\infty } \frac{\exp \left(-|\theta-\theta_0|/\sigma_0\right)}{(x-\theta)^2+\sigma_\epsilon^2} \, d\theta\\
 I_1 &=  \int_{-\infty }^{\infty } \frac{\theta\exp \left(-|\theta-\theta_0|/\sigma_0\right)}{(x-\theta)^2+\sigma_\epsilon^2} \, d\theta.
\end{align*}

Define $u = (\theta-\theta_0)/\sigma_0$, $z_0=x_0/\sigma_0=(x-\theta_0)/\sigma_0$, and $a=\sigma_\epsilon/\sigma_0$. The required normalized integrals are
\begin{align*}
J_0 &= \int_{-\infty }^{\infty } \frac{\exp\paranth{-|u|}}{(u-z_0)^2+a^2}du=\frac{1}{a}\text{Im}(2\pi i\text{cosh}(z)
-\text{Eix}(z) + \text{Eix}(-z))\\
J_1 &= \int_{-\infty }^{\infty } \frac{u\exp\paranth{-|u|}}{(u-z_0)^2+a^2}du= 
\frac{1}{a} \text{Im}\paranth{z\paranth{2\pi i \text{cosh}(z)- \text{Eix}(z) + \text{Eix}(-z)}}
\end{align*}
where $z=z_0+ai$ and $\text{Eix}(\cdot)$ is the scaled exponential integral
\begin{align*}
\text{Eix}(z) = e^{-z}\operatorname{Ei}(z),
\end{align*}
with $\operatorname{Ei}(\cdot)$ taken on its principal branch.

The original integrals become
\begin{align*}
I_0 &= \frac{1}{\sigma_0}J_0\\
I_1 &= \frac{\theta_0}{\sigma_0}J_0 + J_1,
\end{align*}
and the posterior mean becomes
\begin{align*}
    \theta_1 & = \frac{I_1}{I_0}\\
    & = \theta_0  + \sigma_0\frac{J_1}{J_0}\\
    & = \theta_0 + \sigma_0\frac{\text{Im}\paranth{z\paranth{2\pi i \text{cosh}(z)- \text{Eix}(z) +\text{Eix}(-z)}}}{\text{Im}\paranth{2\pi i \text{cosh}(z)- \text{Eix}(z) + \text{Eix}(-z)}}\\
    & = x +\sigma_\epsilon\frac{\text{Re}\paranth{2\pi i \text{cosh}(z)- \text{Eix}(z) +\text{Eix}(-z)}}
    {\text{Im}\paranth{2\pi i \text{cosh}(z)- \text{Eix}(z) +\text{Eix}(-z)}}
\end{align*}

For small signals, $z_0\approx0$, define $A,B\in\mathbb R$ by
\begin{align*}
    A + i B &= \text{Eix}(z) - \text{Eix}(-z)\\
    \frac{d A}{d z_0} + i \frac{d B}{d z_0}&= \frac{d}{d z_0} (\text{Eix}(z) - \text{Eix}(-z))\\
    &= -(\text{Eix}(-z) + \text{Eix}(z))
\end{align*}
which implies
\begin{align*}
    A - \frac{d A}{d z_0}&= 2 \text{Re}(\text{Eix}(z))\\
    B - \frac{d B}{d z_0}&= 2 \text{Im}(\text{Eix}(z))
\end{align*}

Choosing integration paths symmetric about the real axis gives $\text{Eix}(-ia)=\overline{\text{Eix}(ia)}$. Hence, when $z_0\approx0$ and $z\approx ia$,
\begin{align*}
    \text{Eix}(z) - \text{Eix}(-z) &\approx \text{Eix}(i a) - \text{Eix}(-i a)\\
    &= i 2 \text{Im}(\text{Eix}(i a))\\
    \text{Eix}(z) + \text{Eix}(-z) &\approx \text{Eix}(i a) + \text{Eix}(-i a)\\
    &= 2 \text{Re}(\text{Eix}(i a))
\end{align*}
and therefore
\begin{align*}
    A &\approx 0\\
    \frac{d B}{d z_0} &\approx 0\\
    \frac{d A}{d z_0} &\approx  -2 \text{Re}(\text{Eix}(i a))\\    
    B &\approx  2 \text{Im}(\text{Eix}(i a)).
\end{align*}

Define the numerator $N$ and denominator $D$. Near $z_0=0$,
\begin{align*}
    N &= -2\pi \sinh{z_0}\sin{a} - A\\
    &\approx 0 - 0 = 0\\
    D &= 2\pi \cosh{z_0}\cos{a} - B\\
    &\approx 2\pi \cos{a} - 2 \text{Im}(\text{Eix}(i a))\\
    N' &= -2\pi \cosh{z_0}\sin{a} - \frac{d A}{d z_0}\\
    &\approx -2\pi \sin{a} + 2 \text{Re}(\text{Eix}(i a))\\
    D' &= 2\pi \sinh{z_0}\cos{a} - \frac{d B}{d z_0}\\
    &\approx 0 - 0 = 0.
\end{align*}

Differentiating $\theta_1$ with respect to $x$ gives
\begin{align*}
    \omega &= \frac{d}{d x} (x + \sigma_\epsilon \frac{-2\pi \sinh{z_0}\sin{a} - A}{2\pi \cosh{z_0}\cos{a} - B})\\    
    &= 1 + \frac{\sigma_\epsilon}{\sigma_0} \frac{d}{d z_0} \frac{-2\pi \sinh{z_0}\sin{a} - A}{2\pi \cosh{z_0}\cos{a} - B}\\
    &= 1 + a \frac{d}{d z_0} \frac{N}{D}\\
    &= 1 + a (\frac{N'}{D}-\frac{N D'}{D^2})\\ 
    &\approx 1 + a(\frac{-2\pi \sin{a} + 2 \text{Re}(\text{Eix}(i a))}{2\pi \cos{a} - 2 \text{Im}(\text{Eix}(i a))} - 0)\\
    &= 1 + a(\frac{-\pi \sin{a} + \text{Re}(\text{Eix}(i a))}{\pi \cos{a} - \text{Im}(\text{Eix}(i a))})\\    
    &= 1 + a \cdot \frac{-\pi \sin{a} + \cos{a}\text{Ci}(a) + \sin{a}\text{Si}(a) + \frac{\pi}{2}\sin{a}}{\pi \cos{a} - \paranth{\cos{a}\text{Si}(a) + \frac{\pi}{2}\cos{a} - \sin{a}\text{Ci}(a)}}\\
    &= 1 - a \cdot \frac{\frac{\pi}{2} \sin{a} - \cos{a}\text{Ci}(a) - \sin{a}\text{Si}(a)}{\frac{\pi}{2} \cos{a} - \cos{a}\text{Si}(a) + \sin{a}\text{Ci}(a)}
\end{align*}
where we express $\text{Eix}(ia)$ as
\begin{align*}
    \text{Eix}(ia) &= e^{-i a} \operatorname{Ei}(ia)\\
    &= (\cos{a} - i \sin{a}) (\text{Ci}(a) + i\text{Si}(a)+ i \frac{\pi}{2})\\
    &= \cos{a}\text{Ci}(a) + \sin{a}\text{Si}(a) + \frac{\pi}{2}\sin{a} + i \paranth{\cos{a}\text{Si}(a) + \frac{\pi}{2}\cos{a} - \sin{a}\text{Ci}(a)}.
\end{align*}

For $a\gg1$, repeated integration by parts gives
\begin{align*}
    \text{Si}(a) &\approx \frac{\pi}{2} - \frac{\cos a}{a} - \frac{\sin a}{a^2} + \frac{2\cos a}{a^3} + \frac{6 \sin a}{a^4}\\
    \text{Ci}(a) &\approx \frac{\sin a}{a} - \frac{\cos a}{a^2} - \frac{2 \sin a}{a^3} + \frac{6 \cos a}{a^4}
\end{align*}

Substitution into the DeGroot coefficient yields
\begin{align*}
    \omega &\approx 1 - a \cdot \frac{\frac{\pi}{2} \sin{a} - \cos{a}\paranth{\frac{\sin a}{a} - \frac{\cos a}{a^2} - \frac{2 \sin a}{a^3} + \frac{6 \cos a}{a^4}} - \sin{a}\paranth{\frac{\pi}{2} - \frac{\cos a}{a} - \frac{\sin a}{a^2} + \frac{2\cos a}{a^3} + \frac{6 \sin a}{a^4}}}{\frac{\pi}{2} \cos{a} - \cos{a}\paranth{\frac{\pi}{2} - \frac{\cos a}{a} - \frac{\sin a}{a^2} + \frac{2\cos a}{a^3} + \frac{6 \sin a}{a^4}} + \sin{a}\paranth{\frac{\sin a}{a} - \frac{\cos a}{a^2} - \frac{2 \sin a}{a^3} + \frac{6 \cos a}{a^4}}}\\
    &= 1 - a \cdot \frac{\frac{1}{a^2}-\frac{6}{a^4}}{\frac{1}{a}-\frac{2}{a^3}}\\
    &= 1 - \frac{1-\frac{6}{a^2}}{1-\frac{2}{a^2}}\\
    &\approx 1 - (1-\frac{6}{a^2})(1+\frac{2}{a^2})\\
    &= \frac{4}{a^2}.
\end{align*}

For $a \ll 1$ we use
\begin{align*}
    \sin{a} &\approx a\\
    \cos{a} &\approx 1\\
    \text{Si}(a) &\approx a\\
    \text{Ci}(a) &\approx \gamma + \log{a}.
\end{align*}
Consequently, the numerator and denominator in the exact coefficient satisfy
\begin{align*}
    \frac{\pi}{2}\sin a-\cos a\,\text{Ci}(a)-\sin a\,\text{Si}(a)
    &=-\log a-\gamma+O(a),\\
    \frac{\pi}{2}\cos a-\cos a\,\text{Si}(a)+\sin a\,\text{Ci}(a)
    &=\frac{\pi}{2}+O(a|\log a|).
\end{align*}
It follows that
\begin{align*}
    \omega
    =1+\frac{2}{\pi}a\log a+O(a).
\end{align*}

For the large-signal behavior, it is more transparent to expand the Cauchy likelihood directly. Let $U=\Theta-\theta_0$ and $y=x-\theta_0$. Apart from a factor independent of $U$, the likelihood is
\begin{align*}
    L_y(U)=\frac{1}{(y-U)^2+\sigma_\epsilon^2}.
\end{align*}
For large positive $y$,
\begin{align*}
    L_y(U)
    &=\frac{1}{y^2}
    \left(1-\frac{2U}{y}
    +\frac{U^2+\sigma_\epsilon^2}{y^2}\right)^{-1}\\
    &=\frac{1}{y^2}
    \left[1+\frac{2U}{y}
    +\frac{3U^2-\sigma_\epsilon^2}{y^2}
    +O(y^{-3})\right].
\end{align*}
The Laplace prior has exponentially decaying tails, so these terms may be integrated term by term. Since it is centered and symmetric, $\mathbb E[U]=\mathbb E[U^3]=0$. The posterior-mean shift is therefore
\begin{align*}
    \theta_1-\theta_0
    &=\frac{\mathbb E[U L_y(U)]}{\mathbb E[L_y(U)]}\\
    &=\frac{\frac{2\mathbb E[U^2]}{y}+O(y^{-3})}
    {1+\frac{3\mathbb E[U^2]-\sigma_\epsilon^2}{y^2}+O(y^{-4})}\\
    &=\frac{2\operatorname{Var}(\Theta)}{y}+O(y^{-3}).
\end{align*}
For $\Theta\sim\text{Laplace}(\theta_0,\sigma_0)$,
\begin{align*}
    \operatorname{Var}(\Theta)=2\sigma_0^2.
\end{align*}
Consequently,
\begin{align*}
    \theta_1-\theta_0
    &=\frac{4\sigma_0^2}{x-\theta_0}
    +O\left((x-\theta_0)^{-3}\right).
\end{align*}
Thus, for $|x_0|=|x-\theta_0|\gg\sigma_0,\sigma_\epsilon$, the leading approximation is
\begin{align*}
    \theta_1-\theta_0
    \approx\frac{4\sigma_0^2}{x-\theta_0}.
\end{align*}
Odd symmetry supplies the corresponding expansion in the negative tail.

\subsection{Derivations for Localized Response Mechanisms}
\subsubsection{Bimodal Gaussian Prior and Local Overreaction}\label{app:bimodal_overreaction}
Let
\begin{align*}
    \Theta &\sim \tfrac12 N(\theta_0-d,\sigma_0^2)
    +\tfrac12 N(\theta_0+d,\sigma_0^2),
    &\epsilon&\sim N(0,\sigma_\epsilon^2),
\end{align*}
and set $b=0$ and $x_0=x-\theta_0$. Conditional on the negative or positive prior component, Gaussian conjugacy gives posterior component means
\begin{align*}
    \theta_{1,\pm}
    =\theta_0
    +\frac{\sigma_0^2}{\sigma_0^2+\sigma_\epsilon^2}x_0
    \pm\frac{\sigma_\epsilon^2}{\sigma_0^2+\sigma_\epsilon^2}d
\end{align*}
and common posterior variance
\begin{align*}
    \frac{\sigma_0^2\sigma_\epsilon^2}
    {\sigma_0^2+\sigma_\epsilon^2}.
\end{align*}
The posterior component weights, meaning the posterior probabilities that the prior component was the negative or positive one, are
\begin{align*}
    \beta_\pm(x_0)
    =\frac{
    \exp\left[-\frac{(x_0\mp d)^2}{2(\sigma_0^2+\sigma_\epsilon^2)}\right]
    }{
    \exp\left[-\frac{(x_0-d)^2}{2(\sigma_0^2+\sigma_\epsilon^2)}\right]
    +
    \exp\left[-\frac{(x_0+d)^2}{2(\sigma_0^2+\sigma_\epsilon^2)}\right]
    },
\end{align*}
so that
\begin{align*}
    \beta_+(x_0)-\beta_-(x_0)
    =\tanh\left(\frac{d x_0}{\sigma_0^2+\sigma_\epsilon^2}\right).
\end{align*}
The hyperbolic-tangent form is the centered-logistic representation of the posterior odds between the two Gaussian components.
Combining the two components, $\theta_1=\beta_-(x_0)\theta_{1,-}+\beta_+(x_0)\theta_{1,+}$, gives
\begin{align*}
    \theta_1
    =\theta_0
    +\frac{\sigma_0^2}{\sigma_0^2+\sigma_\epsilon^2}x_0
    +\frac{\sigma_\epsilon^2d}{\sigma_0^2+\sigma_\epsilon^2}
    \tanh\left(\frac{d x_0}{\sigma_0^2+\sigma_\epsilon^2}\right).
\end{align*}
Differentiation with respect to $x$ yields
\begin{align*}
    \frac{d\theta_1}{dx}
    =\frac{\sigma_0^2}{\sigma_0^2+\sigma_\epsilon^2}
    +\frac{d^2\sigma_\epsilon^2}{(\sigma_0^2+\sigma_\epsilon^2)^2}
    \operatorname{sech}^2\left(\frac{d x_0}{\sigma_0^2+\sigma_\epsilon^2}\right).
\end{align*}
At $x_0=0$, the posterior mean equals $\theta_0$, while the local coefficient is
\begin{align*}
    \omega_0
    =\frac{\sigma_0^2}{\sigma_0^2+\sigma_\epsilon^2}
    +\frac{d^2\sigma_\epsilon^2}{(\sigma_0^2+\sigma_\epsilon^2)^2}.
\end{align*}
Since
\begin{align*}
    \omega_0-1
    =\frac{\sigma_\epsilon^2
    (d^2-\sigma_0^2-\sigma_\epsilon^2)}
    {(\sigma_0^2+\sigma_\epsilon^2)^2},
\end{align*}
it follows that $\omega_0>1$ if and only if $d^2>\sigma_0^2+\sigma_\epsilon^2$. Moreover,
\begin{align*}
    \frac{\theta_1-\theta_0}{x_0}
    \longrightarrow
    \frac{\sigma_0^2}{\sigma_0^2+\sigma_\epsilon^2}<1
    \qquad\text{as }|x_0|\to\infty,
\end{align*}
so the overreaction is local rather than asymptotic.

\subsubsection{General Finite-Mixture Prior}\label{app:mixture_prior}
Let the prior be
\begin{align*}
    f_\Theta(\theta)=\sum_{i=1}^k\pi_i f_i(\theta),
    \qquad
    \pi_i\geq0,
    \qquad
    \sum_{i=1}^k\pi_i=1,
\end{align*}
where component $i$ has finite mean $\mu_i$. The prior mean is
\begin{align*}
    \theta_0=\sum_{i=1}^k\pi_i\mu_i.
\end{align*}
For the perceived-signal likelihood $l_\epsilon(x-b-\theta)$, define the component marginal signal density and its first unnormalized posterior moment by
\begin{align*}
    J_{0,i}(x)
    &=\int_{-\infty}^{\infty}
    f_i(\theta)l_\epsilon(x-b-\theta)\,d\theta,\\
    J_{1,i}(x)
    &=\int_{-\infty}^{\infty}
    \theta f_i(\theta)l_\epsilon(x-b-\theta)\,d\theta.
\end{align*}
The posterior mean conditional on component $i$ is
\begin{align*}
    \theta_{1,i}(x)=\frac{J_{1,i}(x)}{J_{0,i}(x)},
\end{align*}
and Bayes' rule gives its posterior component weight as
\begin{align*}
    \beta_i(x)
    =\frac{\pi_iJ_{0,i}(x)}
    {\sum_{j=1}^k\pi_jJ_{0,j}(x)}.
\end{align*}
The full posterior mean is therefore
\begin{align*}
    \theta_1
    &=\frac{\sum_{i=1}^k\pi_iJ_{1,i}(x)}
    {\sum_{i=1}^k\pi_iJ_{0,i}(x)}\\
    &=\sum_{i=1}^k\beta_i(x)\theta_{1,i}(x).
\end{align*}
Equivalently, its shift from the prior mean decomposes as
\begin{align*}
    \theta_1-\theta_0
    &=\sum_{i=1}^k\beta_i(x)
    \bigl(\theta_{1,i}(x)-\mu_i\bigr)
    +\sum_{i=1}^k\bigl(\beta_i(x)-\pi_i\bigr)\mu_i.
\end{align*}
The first term is the posterior-weighted within-component update. The second is the shift caused by reweighting the prior components after observing the signal.

To obtain the general response slope, define
\begin{align*}
    r_i(x)&=\frac{d}{dx}\log J_{0,i}(x),
    &\omega_i(x)&=\frac{d\theta_{1,i}(x)}{dx},
\end{align*}
under the regularity conditions permitting differentiation under the integral. Differentiating the posterior component weight gives
\begin{align*}
    \beta_i'(x)
    =\beta_i(x)
    \left[r_i(x)-\sum_{j=1}^k\beta_j(x)r_j(x)\right].
\end{align*}
Consequently,
\begin{align*}
    \frac{d\theta_1}{dx}
    &=\sum_{i=1}^k\beta_i(x)\omega_i(x)
    +\sum_{i=1}^k\beta_i(x)
    \bigl(\theta_{1,i}(x)-\theta_1(x)\bigr)r_i(x)\\
    &=\sum_{i=1}^k\beta_i(x)\omega_i(x)
    +\operatorname{Cov}_{\beta(x)}\!\left(\theta_{1,i}(x),r_i(x)\right).
\end{align*}
Thus a mixture prior produces local overreaction at a signal value $x$ whenever
\begin{align*}
    \sum_{i=1}^k\beta_i(x)\omega_i(x)
    +\operatorname{Cov}_{\beta(x)}\!\left(\theta_{1,i}(x),r_i(x)\right)>1.
\end{align*}
The symmetric two-component Gaussian mixture in Appendix~\ref{app:bimodal_overreaction} is a closed-form example in which the positive covariance generated by component reweighting is large enough to satisfy this inequality.


\subsubsection{Mixture Signal}
Define the unnormalized posterior moments
\begin{align*}
    I_0 & = \phi_\epsilon \int_{-\infty}^{\infty} f_\Theta(\theta) l_\epsilon(x-\theta) d\theta + \sum_{i=1}^k\phi_i\int_{-\infty}^{\infty} f_\Theta(\theta) l_i(x-\Delta_i) d\theta\\
    &= \phi_\epsilon J_0 +  \sum_{i=1}^k\phi_i l_i(x-\Delta_i)\\
    I_1 & = \phi_\epsilon \int_{-\infty}^{\infty} \theta  f_\Theta(\theta) l_\epsilon(x-\theta) d\theta + 
    \sum_{i=1}^k \phi_i \int_{-\infty}^{\infty} \theta f_\Theta(\theta) l_i(x-\Delta_i) d\theta\\
    &=\phi_\epsilon J_1 + \theta_0  \sum_{i=1}^k \phi_i l_i(x-\Delta_i)
\end{align*}
where
\begin{align*}
    J_0 &= \int_{-\infty}^{\infty} f_\Theta(\theta) l_\epsilon(x-\theta) d\theta\\
    J_1 &= \int_{-\infty}^{\infty} \theta f_\Theta(\theta) l_\epsilon(x-\theta) d\theta.
\end{align*}
Here the prior mean is $\theta_0=\int_{-\infty}^{\infty}\theta f_\Theta(\theta)d\theta$. Let $\theta_1^{nm}=J_1/J_0$ denote the posterior mean under the non-mixture signal. The posterior mean under the mixture is
\begin{align*}
    \theta_1 &= \frac{I_1}{I_0}\\
             &= \frac{\phi_\epsilon J_1 + \theta_0  \sum_{i=1}^k \phi_i l_i(x-\Delta_i)}{\phi_\epsilon J_0 + \sum_{i=1}^k\phi_i l_i(x-\Delta_i)}\\
             &= \frac{\theta_1^{nm} \phi_\epsilon + \theta_0  \sum_{i=1}^k \phi_i l_i(x-\Delta_i)/J_0}{\phi_\epsilon + \sum_{i=1}^k\phi_i l_i(x-\Delta_i)/J_0}\\
             &= \alpha \theta_1^{nm} + (1-\alpha) \theta_0,
\end{align*}
where
\[
\alpha = \frac{\phi_\epsilon J_0}{\phi_\epsilon J_0 + \sum_{i=1}^k\phi_i l_i(x-\Delta_i)} \in [0,1].
\]
Thus,
\[
\theta_1 - \theta_0 = \alpha (\theta_1^{nm} - \theta_0) \propto \alpha.
\]
Because $\alpha$ depends on $x$, the degree of attenuation varies across observations. When $x$ is more likely to come from the $\Theta$-centered component, $\alpha$ is larger and the shift is greater. Near a state-independent component $\Delta_i$, $\alpha$ is smaller, producing localized attenuation.

\subsubsection*{Gaussian Prior, Gaussian Mixture Signal}
For the special case of a prior centered at zero, so that $\theta_0=0$,
\begin{align*}
    \theta_1 &= \alpha \theta_1^{nm}\\
    &= \theta_1^{nm} \paranth{1 +\sum_{i=1}^k \frac{\phi_i}{\phi_\epsilon}\frac{l_i(x-\Delta_i)}{J_0} }^{-1}.
\end{align*}
If the prior is Gaussian and the signal is a Gaussian mixture, the posterior mean is
\begin{align*}
    \theta_1 & = \omega_{nm} x
    \paranth{1 + \sum_{i=1}^k \frac{\phi_i}{\phi_\epsilon} \frac{\sqrt{\sigma_0^2 + \sigma_\epsilon^2}}{\sigma_i} \exp({\frac{x^2}{2(\sigma_0^2 + \sigma_\epsilon^2)} -\frac{(x-\Delta_i)^2}{2\sigma_i^2} })}^{-1}
\end{align*}
where $\omega_{nm} = \frac{\sigma_0^2}{\sigma_0^2 + \sigma_\epsilon^2}$.

Its behavior is governed by the exponent
\begin{align*}
    {\frac{x^2}{2(\sigma_0^2 + \sigma_\epsilon^2)} -\frac{(x-\Delta_i)^2}{2\sigma_i^2} } 
    &= \frac{1}{2(\sigma_0^2 + \sigma_\epsilon^2) \sigma_i^2} [(\sigma_i^2 - \sigma_0^2 - \sigma_\epsilon^2)x^2 + 2 \Delta_i(\sigma_0^2 + \sigma_\epsilon^2) x - (\sigma_0^2 + \sigma_\epsilon^2)\Delta_i^2]\\
    &= \frac{\sigma_i^2 - \sigma_0^2 - \sigma_\epsilon^2}{2(\sigma_0^2 + \sigma_\epsilon^2) \sigma_i^2}x^2 + \frac{\Delta_i}{\sigma_i^2}x - \frac{\Delta_i^2}{2 \sigma_i^2}.
\end{align*}

Near $x=0$, the exponent is approximately $-\Delta_i^2/(2\sigma_i^2)$. The resulting local DeGroot coefficient is
\begin{align*}
    \theta_1 \approx \frac{\omega_{nm}}{1 + \sum_{i=1}^k \frac{\phi_i}{\phi_\epsilon} \frac{\sqrt{\sigma_0^2 + \sigma_\epsilon^2}}{\sigma_i} \exp(-\frac{\Delta_i^2}{2\sigma_i^2})} x.
\end{align*}
If $|\Delta_i|\gg\sigma_i$ for every $i$, so that all state-independent centers are far from the prior belief relative to their scales, we recover the non-mixture DeGroot model:
\begin{align*}
    \theta_1 \approx \omega_{nm} x.
\end{align*}

For large $|x|$, three cases arise according to the sign of $\sigma_i^2-(\sigma_0^2+\sigma_\epsilon^2)$, which compares the state-independent component variance with the variance of the Gaussian convolution.

\begin{enumerate}
    \item If $\sigma_i^2 > \sigma_0^2 + \sigma_\epsilon^2$ for every $i$, the exponent is positive and the opinion shift decays toward zero, producing tail rejection. Although the response is asymmetric, the decay occurs in both tails.
    
    \item If $\sigma_i^2 = \sigma_0^2 + \sigma_\epsilon^2$ for every $i$, the exponent is positive when $x$ and $\Delta_i$ have the same sign, and the shift decays toward zero. When their signs differ, the exponent is negative and the shift follows the DeGroot model.
    
    \item If $\sigma_i^2 < \sigma_0^2 + \sigma_\epsilon^2$ for every $i$, the exponent is negative in the tails and the response returns to the non-mixture DeGroot model:
    \begin{align*}
        \theta_1 \approx \omega_{nm} x.
    \end{align*}
    In an intermediate region, however, the exponent is positive and the shift is smaller than its non-mixture counterpart. This occurs when
    \begin{align*}
        \frac{\sigma_i^2 - \sigma_0^2 - \sigma_\epsilon^2}{2(\sigma_0^2 + \sigma_\epsilon^2) \sigma_i^2}x^2 + \frac{\Delta_i}{\sigma_i^2}x - \frac{\Delta_i^2}{2 \sigma_i^2} > 0.
    \end{align*}
    Solving this inequality for $x$ gives
    \begin{align*}
        x \in
        \begin{cases}
        \left( \Delta_i \frac{1-\sqrt{s}}{1-s}, \; \Delta_i \frac{1+\sqrt{s}}{1-s} \right) & \text{if } \Delta_i > 0, \\[1ex]
        \left( \Delta_i \frac{1+\sqrt{s}}{1-s}, \; \Delta_i \frac{1-\sqrt{s}}{1-s} \right) & \text{if } \Delta_i < 0.
        \end{cases}        
    \end{align*}
    where $s = \frac{\sigma_i^2}{\sigma_0^2+\sigma_\epsilon^2}<1$.
    
    In this regime, the shift is strongly attenuated if $\Delta_i^2 \gg 2 (\sigma_0^2 + \sigma_\epsilon^2 - \sigma_i^2)$. This follows by requiring the quadratic function at $x=\frac{\Delta_i}{1-s}$ to satisfy
    \begin{align*}
        \frac{s-1}{2 \sigma_i^2}(\frac{\Delta_i}{1-s})^2 + \frac{\Delta_i}{\sigma_i^2}(\frac{\Delta_i}{1-s}) - \frac{\Delta_i^2}{2 \sigma_i^2} \gg 0.
    \end{align*}
    
    In the special case $\sigma_i = \sigma_\epsilon$, so that $s=1-\omega_{nm}$, the intermediate regime becomes
    \begin{align*}
        x \in
        \begin{cases}
        \left(\dfrac{\Delta_i}{\omega_{nm}}(1 - \sqrt{1 - \omega_{nm}}),\; \dfrac{\Delta_i}{\omega_{nm}}(1 + \sqrt{1 - \omega_{nm}})\right) & \text{if } \Delta_i > 0, \\[1ex]
        \left(\dfrac{\Delta_i}{\omega_{nm}}(1 + \sqrt{1 - \omega_{nm}}),\; \dfrac{\Delta_i}{\omega_{nm}}(1 - \sqrt{1 - \omega_{nm}})\right) & \text{if } \Delta_i < 0.
        \end{cases}
    \end{align*}
\end{enumerate}

\subsubsection*{Gaussian Prior, Gaussian Mixture Signal with Uncertain Bias Points and Weights}
Suppose the receiver holds a Gaussian prior over the latent state, $\Theta \sim N(\theta_0, \sigma_0^2)$. The perceived-signal likelihood is a mixture of an informative Gaussian component centered at $\Theta$ and $k$ state-independent Gaussian components with unknown centers and uncertain weights. Assume that $\Theta$, the component locations, and the random weight vector $\boldsymbol\Phi$ are mutually independent. Conditional on $\boldsymbol\Phi$, the likelihood is
\[
l_{X|\Theta,\boldsymbol\Phi}(x|\theta,\boldsymbol\phi)
= \phi_\epsilon N(x; \theta, \sigma_\epsilon^2)
+ \sum_{i=1}^k \phi_i l_i(x).
\]

Each state-independent signal component $l_i(x)$ is generated in two stages. First, its latent location is drawn from $\Delta_i \sim N(\delta_i, \tau_i^2)$. Conditional on $\Delta_i$ and component $i$, the signal is Gaussian:
\[
X\mid\Delta_i,i \sim N(\Delta_i, \sigma_i^2).
\]
Marginalizing over the uncertain location gives
\[
l_i(x) = \int N(x; \Delta_i, \sigma_i^2) \cdot N(\Delta_i; \delta_i, \tau_i^2) \, d\Delta_i.
\]
This Gaussian convolution yields
\[
l_i(x) = N(x; \delta_i, \tau_i^2 + \sigma_i^2).
\]

Integrating over the component locations and random weights gives the effective likelihood
\[
l_{X|\Theta}(x|\theta)
= \bar\phi_\epsilon N(x; \theta, \sigma_\epsilon^2)
+ \sum_{i=1}^k \bar\phi_i N(x; \delta_i, \tau_i^2 + \sigma_i^2),
\]
where the random weights have the Dirichlet prior
\[
(\Phi_\epsilon, \Phi_1, \dots, \Phi_k) \sim \text{Dir}(\alpha_\epsilon, \alpha_1, \dots, \alpha_k).
\]
Under this prior, the expected weight of component $j$ is
\[
\bar{\phi}_j \triangleq
\mathbb{E}[\Phi_j] = \frac{\alpha_j}{\sum_{i=1}^k \alpha_i + \alpha_\epsilon}.
\]
This expectation holds for both the informative component ($j = \epsilon$) and each state-independent component ($j = 1, \dots, k$).

The informative likelihood component is conjugate to the Gaussian prior and yields the non-mixture posterior mean (see Section~\ref{sec:DeGroot}):
\[
\theta_1^{nm} = \left( \frac{\theta_0}{\sigma_0^2} + \frac{x}{\sigma_\epsilon^2} \right) \Big/ \left( \frac{1}{\sigma_0^2} + \frac{1}{\sigma_\epsilon^2} \right).
\]

Although the full posterior is not Gaussian, its mean remains a convex combination of the prior mean $\theta_0$ and the informative posterior mean $\theta_1^{nm}$:
\[
\theta_1 = \alpha(x) \cdot \theta_1^{nm} + (1 - \alpha(x)) \cdot \theta_0,
\]
where the weight $\alpha(x)$ reflects the posterior belief that $x$ originated from the informative component:
\[
\alpha(x) = \frac{\bar{\phi}_\epsilon \cdot N(x; \theta_0, \sigma_0^2 + \sigma_\epsilon^2)}{\bar{\phi}_\epsilon \cdot N(x; \theta_0, \sigma_0^2 + \sigma_\epsilon^2) + \sum_{i=1}^k \bar{\phi}_i \cdot N(x; \delta_i, \tau_i^2 + \sigma_i^2)} \in [0,1].
\]

Thus the posterior mean retains a tractable convex form even when the state-independent components have uncertain locations and weights. The receiver updates substantially only when $x$ is relatively more likely to have come from the informative component.

\end{document}